\documentclass{aa}
\usepackage{natbib}
\usepackage{aalongtable}
\usepackage{graphics}
\usepackage{epsfig}
\usepackage{color}
\usepackage{times}
\newcommand\ra[4]{~~#1$^{\rm h}$#2$^{\rm m}$#3$^{\rm s}$.#4 }
\newcommand\dec[3] {$#1$$^{\circ}$#2$^{\rm '}$#3$^{\rm ''}$}

\newcommand{\sbu}{mag arcsec$^{-2}$}
\newcommand{\sbb}{mag/$\sq\arcsec$}
\def\ha{H$\alpha$}
\def\h2{H{\small~II}}
\def\lowspace{\rule[-1.25ex]{0cm}{1.25ex}}
\def\upperspace{\rule[0.0ex]{0cm}{2.5ex}}
\def\apj{ApJ}
\def\apjs{ApJS}
\def\aap{A\&A}
\def\aaps{A\&AS}
\def\aj{AJ}
\def\mnras{MNRAS}
\newfont{\vs}{cmssdc10 scaled 1000}
\def\bb{}
\begin{document}
\title{New southern blue compact dwarf galaxies in the 2dF Galaxy 
Redshift Survey
\thanks{Based on observations collected 
at the European Southern Observatory, Chile, ESO program 71.B-0032.}}
\author{P.\ Papaderos \inst{1}
\and N.G.\ Guseva \inst{2}
\and Y.I.\ Izotov \inst{2}
\and K.G.\ Noeske\inst{1,3}
\and T.X.\ Thuan \inst{4}
\and K.J.\ Fricke \inst{1}}
\offprints{P. Papaderos,\\ papade@astro.physik.uni-goettingen.de}
\institute{     Institut f\"ur Astrophysik, Friedrich-Hund-Platz 1, 
                D--37077 G\"ottingen, Germany
\and
                 Main Astronomical Observatory, 
                 Ukrainian National Academy of Sciences,
                 27 Zabolotnoho str., Kyiv 03680,  Ukraine
\and
                 University of California, 1156 High St., Santa Cruz, 
                 CA 95064, USA
\and
                 Astronomy Department, University of Virginia, 
                 Charlottesville, VA 22903, USA
}

\date{Received \hskip 2cm; Accepted}

\abstract{}{Aiming to find new extremely metal-deficient star-forming galaxies
we extracted from the Two-Degree Field Galaxy Redshift Survey (2dFGRS) 
100K Data Release 14 emission-line galaxies with relatively strong 
[O {\sc iii}] $\lambda$4363 emission.}{Spectroscopic and photometric studies of this sample and, 
in addition, of 7 Tololo and 2 UM galaxies were performed 
on the basis of observations with the ESO 3.6m telescope.}
{All sample galaxies qualify with respect to their photometric 
and spectroscopic properties as blue compact dwarf (BCD) galaxies. 
Additionally, they show a good overlap with a comparison sample 
of $\sim$ 100 well-studied emission-line galaxies on the 
12+log(O/H) vs. log(Ne/O), log(Ar/O) and log(Fe/O) planes.
From the analysis of the 2dFGRS subsample we report the discovery of 
two new extremely metal-deficient BCDs with an oxygen 
abundance 12+log(O/H) $\leq$ 7.6 and of another seven galaxies 
with 12+log(O/H) $\la$ 7.8. 
Furthermore, we confirm previous oxygen abundance determinations 
for the BCDs \object{Tol 1304--353}, \object{Tol 2146--391}, \object{UM 559} and \object{UM 570} 
to be 12+log(O/H) $\leq$ 7.8.}{} 
\maketitle

%
\section {Introduction}\label{intro}
The identification of chemically unevolved star-forming galaxies 
in the local universe, with a nearly pristine chemical composition, 
is a fundamental quest of contemporary observational cosmology. 
This is because of several reasons.

First, studies of abundance ratios and abundance gradients in these systems 
are crucial for our understanding of the early chemical evolution 
of galaxies and for constraining models of stellar nucleosynthesis and 
the properties of massive stars in the early universe. 

Second, if we wish to understand the spectra of primeval galaxies, it is highly important 
to understand how the UV radiation field of these systems changes as metallicity decreases. 
It is well established that the lower the metallicity, the harder is the radiation
from massive stars \citep{Campbell86}. 
As a consequence, strong high-ionization lines are expected in the spectra of 
metal-deficient star-forming galaxies. Indeed, deep spectroscopy has led recently 
to the discovery of high-ionization emission lines, such as 
[Fe {\sc v}] $\lambda$4227 and [Ne {\sc v}] $\lambda$3346, 3426,
in addition to strong He {\sc ii} $\lambda$4686 emission, in a few of the 
most metal-poor star-forming galaxies known in the local universe 
\citep{Fricke01,ICF01,Izotov2004,TI05}.

Third, systematic studies of star-forming galaxies all the way down 
to a strongly subsolar metallicity are indispensable for putting tight observational
constraints to the primordial helium abundance, which constitutes one of the key tests 
of Standard Big Bang Nucleosynthesis. They allow to infer the ratio d$Y$/d$Z$ 
of the mass fraction $Y$ of helium released into the interstellar medium by stars 
relative to the mass fraction $Z$ of heavy elements. This quantity determines 
the slope of the linear regression of $^4$He against O and N and hence directly 
affects the extrapolation of those regression lines to zero metallicity to 
determine the primordial helium abundance 
{\bb \citep[][ and references therein]{PeimpertTP1974,PeimpertTP1976,Pagel92,IT98,IT04a}}. 

Fourth, recent work lends strong observational support to the idea that some
among the most metal-deficient star-forming galaxies known in the local universe 
have formed most of their stellar mass within the last 1 Gyr, hence they qualify as
young galaxy candidates {\bb \citep{Papaderos2002,Guseva03a,IT04b,PPK04}}. 
Evidently, such systems are excellent nearby laboratories for studying at 
high spatial resolution the early chemical and dynamical evolution of galaxies 
and can yield fundamental insight into the process of galaxy formation
in the distant universe. 
In this regard, dedicated searches for metal-poor star-forming galaxies
at low redshifts are of great interest.

Blue compact dwarf (BCD) galaxies constitute ideal nearby laboratories for studying 
active star formation and galaxy evolution under conditions which in many aspects 
may resemble those in distant low-mass protogalaxies. These systems are gas-rich,
low-luminosity ($M_B\ga$--18) objects undergoing an intense burst of star formation 
in a very compact region ($\leq$1 kpc).
Other than originally conjectured (Sargent \& Searle 1970), the majority of BCDs 
are not primordial systems, but evolved dwarf galaxies where starburst activity 
is immersed within an old extended stellar host galaxy 
\citep{LT86,Kunth88,Papaderos96a}. 

With an oxygen abundance 12+log(O/H) ranging between 7.12 and $\sim$8.4 BCDs 
are the most metal-deficient star-forming galaxies known in the universe.
They are thus excellent laboratories for studying star formation and other physical
processes in nearly primordial conditions, similar to those at the time of 
galaxy formation.
While all BCDs show subsolar chemical abundances, it is notoriously difficult
to find extremely metal-deficient (12+log(O/H)$\la$7.6) objects. 
The BCD abundance distribution peaks at 12+log(O/H)$\approx$8 with a sharp 
drop-off at lower values.

One of the first BCDs discovered, I Zw 18 \citep{SS1970} 
with 12+log(O/H) of 7.17$\pm$0.01 in its northwest component and of
7.22$\pm$0.02 in its southeast component \citep{TI05} 
continued to hold the record as the most metal-deficient BCD known for more than 
three decades. Only very recently, has this system displaced in the metallicity
ranking by the BCD SBS 0335--052\,W with an oxygen abundance 
12+log(O/H)=7.12$\pm$0.03 \citep{ITG05}.

So far, less than a couple of ten BCDs with an oxygen abundance 12 + log(O/H) $\leq$ 7.6 
have been discovered despite the existence of a extensive
amount of data from large CCD sky surveys performed with modern telescopes.
For instance, out of $\sim$50 000 Sloan Digital Sky Survey (SDSS) galaxy spectra
\citet{SI03} and \citet{ISGT04} were able to
extract only 310 galaxies with detected [O {\sc iii}] $\lambda$4363 emission, allowing 
for a reliable oxygen abundance determination. 
This sample does not contain any extremely metal-deficient system, 
with 12 + log(O/H) $\leq$ 7.6. 
More recently, \citet{I06} based on a larger sample 
of $\sim$ 500000 galaxies from the SDSS Data Release~3 
also did not find galaxies with 12 + log(O/H) $\leq$ 7.6.

Continuing our search for nearby (redshift $z$ $\la$ 0.1) metal-poor 
star-forming galaxies we focus here on a sample of candidate objects selected from
the Two-Degree Field Galaxy Redshift Survey \citep[2dFGRS ][]{C01}. 
By visual inspection of the 2dFGRS 100K Data Release, we have extracted
$\sim$250 galaxy spectra with strong emission lines and blue continua. 
Out of this sample we have chosen 14 most promising metal-poor galaxy
candidates with strong [O {\sc iii}] $\lambda$4363 emission
for follow-up observations. This is because the latter auroral 
line is easily detectable in high-temperature and hence low-metallicity 
objects. Furthermore, this 
line can be used as a temperature diagnostic, allowing for a direct 
determination of the physical conditions and chemical composition of 
the ionized interstellar medium in galaxies.
This criterion also picks out AGNs, which we have rejected by visual
examination of each spectrum based on the specific relative intensities
of some emission lines (e.g., strong [O {\sc ii}] $\lambda$3727 and
[N {\sc ii}] $\lambda$6583 emission lines in combination with strong
[O {\sc iii}] $\lambda$5007 emission line, strong He {\sc ii} $\lambda$4686
and [O {\sc i}] $\lambda$6300, etc.). 

In this paper we present spectroscopic and photometric follow-up observations 
of the 14 selected 2dFGRS objects. In addition, we include in our study 7 Tololo 
and 2 University of Michigan (UM) Survey emission-line galaxies 
\citep{salzer89a,salzer89b,terl91}. 
The latter objects were included with the purpose of a) improving 
their abundance determinations and b) enhancing the efficiency of our 
observing programme, filling the gap in right ascension between 
13$^{\rm h}$ and 22$^{\rm h}$ which was not covered by the galaxies from the 2dFGRS.

This paper is organized as follows: the data acquisition 
and reduction are described in Sect. \ref{odr}. 
Our results, based on the spectroscopic and photometric analysis, are presented 
in Sects. \ref{res} and \ref{photo}. 
We summarize our conclusions in Sect. \ref{sum}.

 \section{Observations and data reduction \label{odr}}
%
 \subsection{Spectroscopy \label{spec}}
%
Long slit spectra of the sample galaxies in the 
wavelength range $\lambda$$\lambda$3400 -- 7400 were obtained with 
EFOSC2 (ESO Faint Object Spectrograph and Camera) mounted at the 3.6m ESO 
telescope at La Silla. The programme was carried out in two separate 3-night
observing runs, during the periods April 23-25, 2003 and 
September 30- October 2, 2003. The observing conditions were photometric 
during two nights of the first run (April 24 and 25) but were
subject to large sky transparency variations for the remaining nights.

The journal of observations is given in Table~\ref{obs} with the
target galaxies listed in order of increasing right ascension.
All observations were performed with the same instrumental setup. 
We used grism $\#11$ ($\lambda$$\lambda$3400--7400) with grating 300 gr/mm 
and no second-order blocking filter. The long slit with a width of 1\arcsec\ 
was centered on the brightest H~{\sc ii} region of each galaxy.
The spectra were binned along the spatial and dispersion axes resulting in 
a spatial scale along the slit of 0\farcs314 pixel$^{-1}$, and a
spectral resolution of $\sim$13.2~\AA\ (FWHM).

Our sample galaxies were mostly observed at low airmass 
($<$1.2, cf. Table~\ref{obs}). Spectroscopic observations 
at a higher airmass were carried out along the parallactic angle.
Thus, no corrections for atmospheric refraction have been applied. 
The total exposure time of typically 1 hour per target was split up 
into several subexposures (2 -- 6) to allow for a more 
efficient cosmic ray rejection. 
Spectrophotometric standard stars were observed for flux calibration. 
The seeing varied between $\sim$0\farcs9 and $\sim$1\farcs6.
In Fig.~\ref{Slit} we show the slit orientation for each object superimposed 
on a broad-band CCD V image taken during the same run.


\begin{table*}[tbh]
\caption{List of observed galaxies}
\label{obs}
\begin{tabular}{lccclcccc} \hline \hline

Galaxy &2dFGRS & ($\alpha,\delta$) &  Date &  Exp. time      &  Airmass   & seeing & redshift  \\ 
       & name  &        epoch 2000 &       &  (sec)      &            &  &        \\ \hline
 
\object{2dF 336246}$^{\rm a}$ &S893Z604&\ra{00}{23}{05}{9},\dec{-40}{59}{01}& 30.09.2003 &   3$\,\times\,$1200   & 1.075$^{\rm b}$  & 1\farcs15 & 0.00577 $\pm$ 0.00004 \upperspace \\
\object{2dF 084585}$^{\rm a}$ &   ...  &~~01~~36~~~14.~3,~~~--29~~42~~43    & 02.10.2003 &   3$\,\times\,$1200   & 1.001            & 1\farcs02 & 0.01750 $\pm$ 0.00009  \\
\object{2dF 169299}$^{\rm a}$ &   ...  &~~03~~13~~~02.~7,~~~--26~~14~~14    & 30.09.2003 &   3$\,\times\,$1200   & 1.006            & 0\farcs98 & 0.03841 $\pm$ 0.00011  \\
\object{2dF 366973}$^{\rm a}$ &S909Z481&~~05~~15~~~20.~5,~~~--31~~25~~25    & 30.09.2003 &   2$\,\times\,$1200   & 1.002            & 1\farcs19 & 0.00534 $\pm$ 0.00014  \\
\object{2dF 211354}$^{\rm a}$ &N105Z061&~~11~~01~~~06.~4,~~~--05~~34~~45    & 24.04.2003 &   3$\,\times\,$1200   & 1.327$^{\rm b}$  & 1\farcs26 & 0.02375 $\pm$ 0.00003  \\
\object{2dF 214945}$^{\rm a}$ &N111Z260&~~11~~26~~~55.~5,~~~--04~~47~~54    & 26.04.2003 &   3$\,\times\,$600    & 1.164            & 1\farcs57 & 0.03407 $\pm$ 0.00011  \\
\object{2dF 301921}$^{\rm a}$ &N247Z167&~~12~~13~~~39.~1,~~~--01~~17~~40    & 25.04.2003 &   3$\,\times\,$1200   & 1.137$^{\rm b}$  & 1\farcs00 & 0.00769 $\pm$ 0.00011  \\
\object{2dF 302461}$^{\rm a}$ &N248Z184&~~12~~21~~~55.~7,~~~--01~~35~~20    & 26.04.2003 &   6$\,\times\,$600    & 1.131$^{\rm b}$  & 1\farcs33 & 0.00574 $\pm$ 0.00005  \\
\object{Tol 1304--386}        &   ...  &~~13~~07~~~21.~4,~~~--38~~54~~54    & 24.04.2003 &   3$\,\times\,$900    & 1.016            & 1\farcs49 & 0.01396 $\pm$ 0.00011  \\
\object{Tol 1304--353}        &   ...  &~~13~~07~~~37.~9,~~~--35~~38~~46    & 26.04.2003 &   3$\,\times\,$300    & 1.477$^{\rm b}$  & 1\farcs55 & 0.01387 $\pm$ 0.00012  \\
\object{UM 559}          &   ...  &~~13~~17~~~44.~4,~~~--00~~59~~54    & 26.04.2003 &   3$\,\times\,$900    & 1.176            & 1\farcs57 & 0.00402 $\pm$ 0.00017  \\
\object{UM 570}          &   ...  &~~13~~23~~~47.~5,~~~--01~~32~~59    & 26.04.2003 &   3$\,\times\,$900    & 1.287$^{\rm b}$  & 1\farcs32 & 0.02221 $\pm$ 0.00003  \\
\object{Tol 1334--326}        &   ...  &~~13~~37~~~08.~6,~~~--32~~55~~19    & 24.04.2003 &   3$\,\times\,$900    & 1.020            & 1\farcs36 & 0.01217 $\pm$ 0.00008  \\
\object{2dF 323427}$^{\rm a}$ &N339Z096&~~14~~01~~~07.~5,~~~--00~~41~~01    & 25.04.2003 &   3$\,\times\,$1200   & 1.347            & 1\farcs07 & 0.01213 $\pm$ 0.00009  \\ 
\object{Tol 1400--411}        &   ...  &~~14~~03~~~19.~0,~~~--41~~22~~50    & 24.04.2003 &   3$\,\times\,$600    & 1.090            & 1\farcs00 & 0.00191 $\pm$ 0.00020  \\
\object{Tol 1924--416}        &   ...  &~~19~~27~~~59.~3,~~~--41~~34~~36    & 24.04.2003 &   3$\,\times\,$900    & 1.151            & 1\farcs36 & 0.00951 $\pm$ 0.00012  \\
\object{Tol 2138--405}        &   ...  &~~21~~41~~~21.~8,~~~--40~~19~~06    & 24.04.2003 &   3$\,\times\,$1800   & 1.199            & 1\farcs48 & 0.05805 $\pm$ 0.00028  \\
\object{Tol 2146--391}        &   ...  &~~21~~49~~~48.~9,~~~--38~~54~~25    & 25.04.2003 &   3$\,\times\,$1200   & 1.588$^{\rm b}$  & 1\farcs05 & 0.02936 $\pm$ 0.00011  \\
\object{2dF 042049}$^{\rm a}$ &S479Z167&~~22~~10~~~12.~7,~~~--33~~39~~47    & 01.10.2003 &   4$\,\times\,$1200   & 1.003            & 1\farcs10 & 0.03153 $\pm$ 0.00011  \\  
\object{2dF 171716}$^{\rm a}$ &   ...  &~~22~~13~~~26.~1,~~~--25~~26~~43    & 30.09.2003 &   3$\,\times\,$1200   & 1.067            & 0\farcs76 & 0.01089 $\pm$ 0.00019  \\
\object{2dF 116230}$^{\rm a}$ &S180Z125&~~22~~29~~~01.~5,~~~--27~~51~~05    & 30.09.2003 &   3$\,\times\,$1200   & 1.071            & 1\farcs02 & 0.02332 $\pm$ 0.00016  \\
\object{2dF 181442}$^{\rm a}$ &   ...  &~~22~~36~~~10.~3,~~~--25~~11~~14    & 01.10.2003 &   3$\,\times\,$1200   & 1.065            & 1\farcs10 & 0.01175 $\pm$ 0.00014  \\
\object{2dF 115901}$^{\rm a}$ &S258Z058&~~22~~37~~~03.~1,~~~--28~~52~~22    & 30.09.2003 &   3$\,\times\,$1200   & 1.001            & 0\farcs87 &0.03996 $\pm$ 0.00008  \\
\hline\hline
\upperspace
\end{tabular}

$^{\rm a}$ internal number from the 2dF Galaxy Redshift Survey 100K 
Data Release. \\
$^{\rm b}$ spectra taken along the parallactic angle\\
\end{table*}

The data reduction was carried out with the IRAF\footnote{IRAF is 
the Image Reduction and Analysis Facility distributed by the 
National Optical Astronomy Observatory, which is operated by the 
Association of Universities for Research in Astronomy (AURA) under 
cooperative agreement with the National Science Foundation (NSF).}
software package. 
This includes bias--subtraction, flat--field correction, cosmic-ray removal, 
wavelength calibration, night sky background subtraction, correction for 
atmospheric extinction and absolute flux calibration of the two--dimensional spectrum.

Redshift-corrected one-dimensional spectra of the brightest H {\sc ii} region in each 
galaxy (labeled {\tt a} in Fig. \ref{Slit}) are shown in Fig.~\ref{spectra1}.
Additionally, we extracted spectra for several of the fainter H\,{\sc ii} regions 
labeled in Fig. \ref{Slit}. A summary of the spectroscopic properties
of the H {\sc ii} regions is given in Table~\ref{t:HIIadd}. 

\begin{figure*}[th]
\begin{picture}(24,19)
\put(0.0,0.0){{\psfig{figure=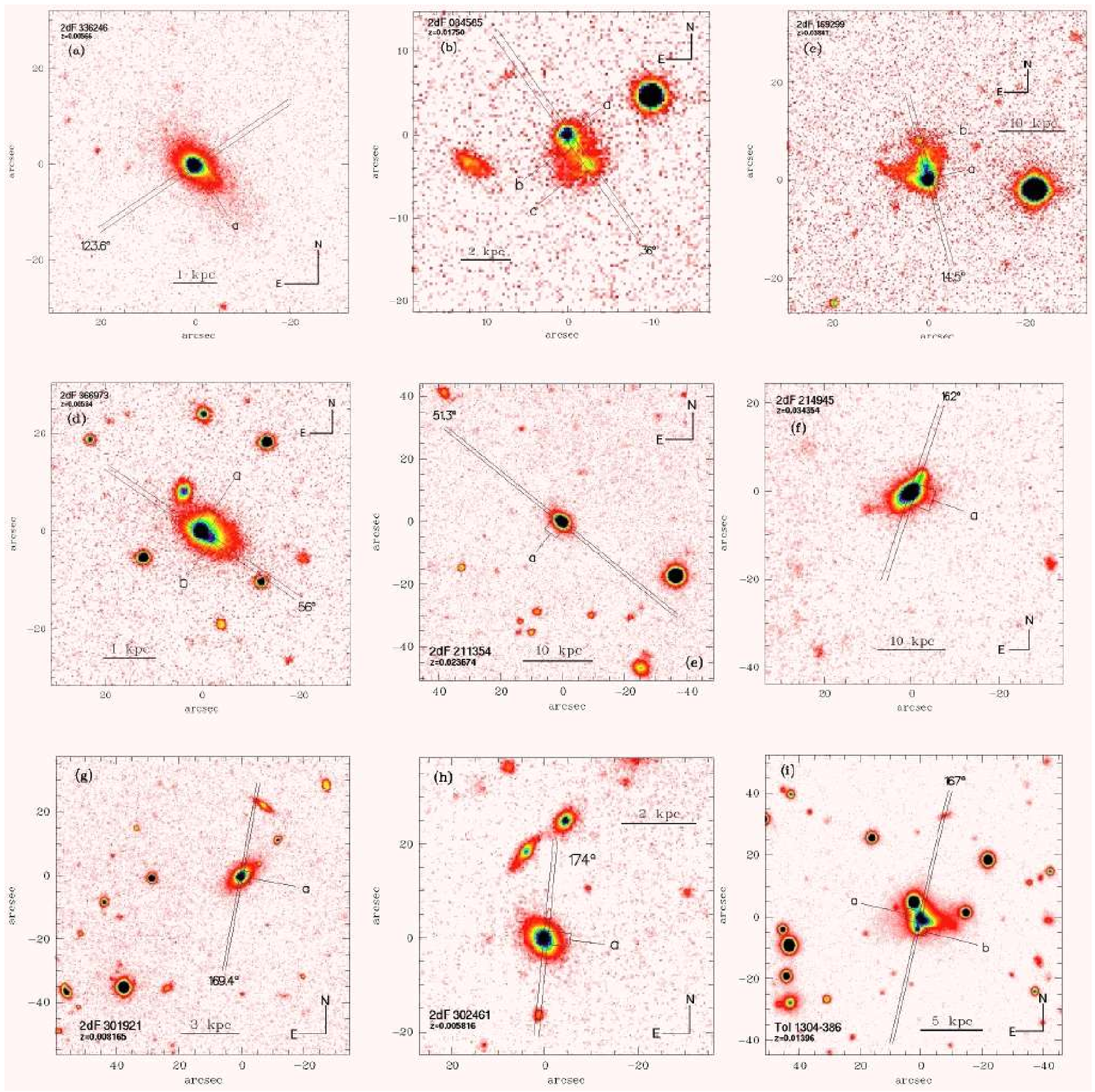,height=18.5cm,angle=0.,clip=}}}
\end{picture}
\caption{Reproduction of the slit position on $V$ band EFOSC2 images of our
sample galaxies. For \object{Tol 1400--411} (panel o) for which no $V$ image
was acquired we use instead a $J$ band exposure from
\citet{Kai2003}. Selected {\bb \h2} regions from which 1D spectra were extracted 
are indicated. Note that the faint arc-like features in panels {\sl j} (\object{Tol 1304--353}) and 
{\sl m} (\object{Tol 1334--326}) are artefacts due to scattered light from nearby bright stars. 
In all panels north is up and east to the left.}
\label{Slit}
\end{figure*}

\setcounter{figure}{0}
\begin{figure*}
\begin{picture}(24,19)
\put(0.0,0.0){{\psfig{figure=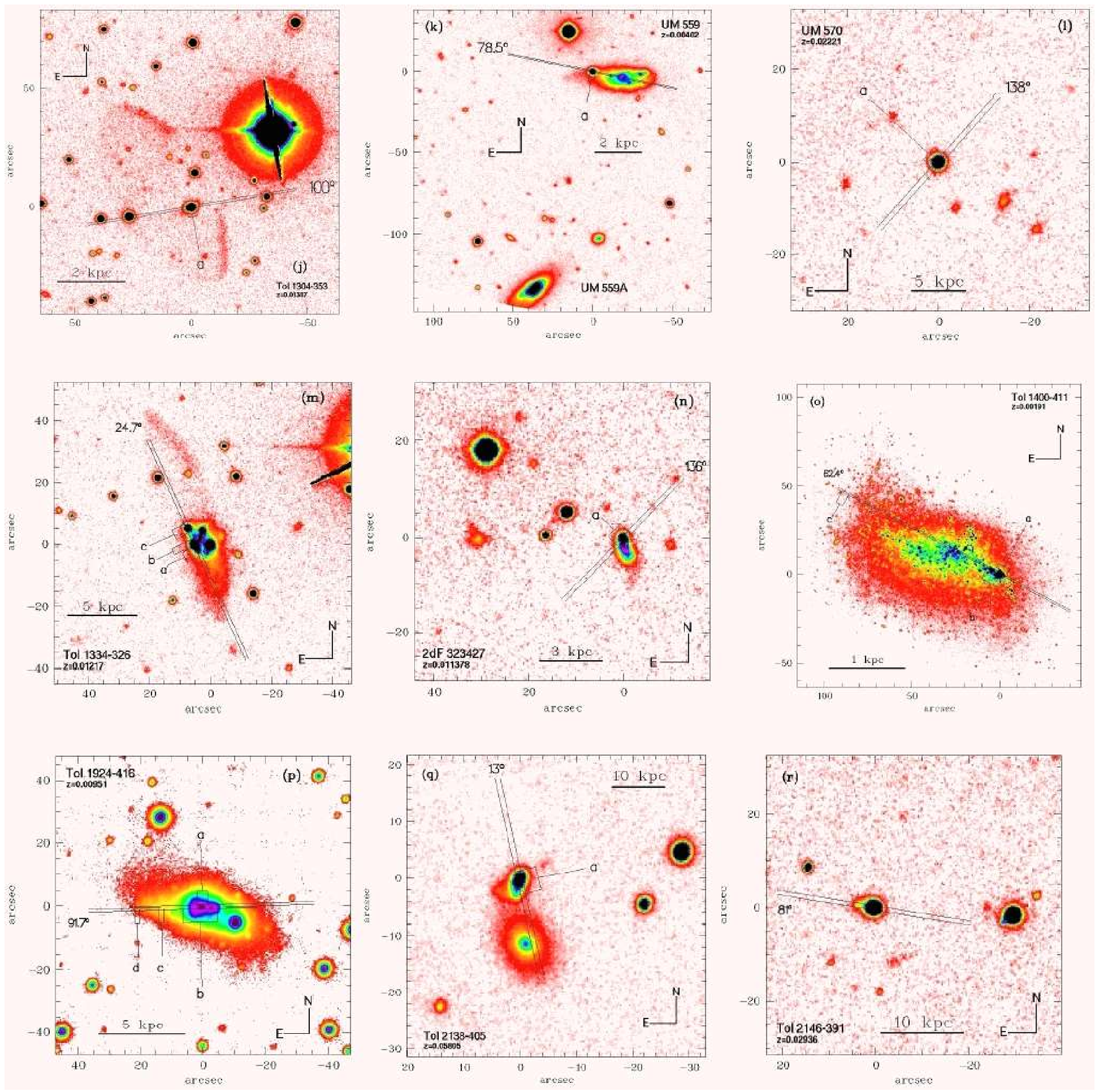,height=18.5cm,angle=0.,clip=}}}
\end{picture}
\caption{continued}
\end{figure*}

\setcounter{figure}{0}
\begin{figure*}
\begin{picture}(24,12.5)
\put(0.0,0.0){{\psfig{figure=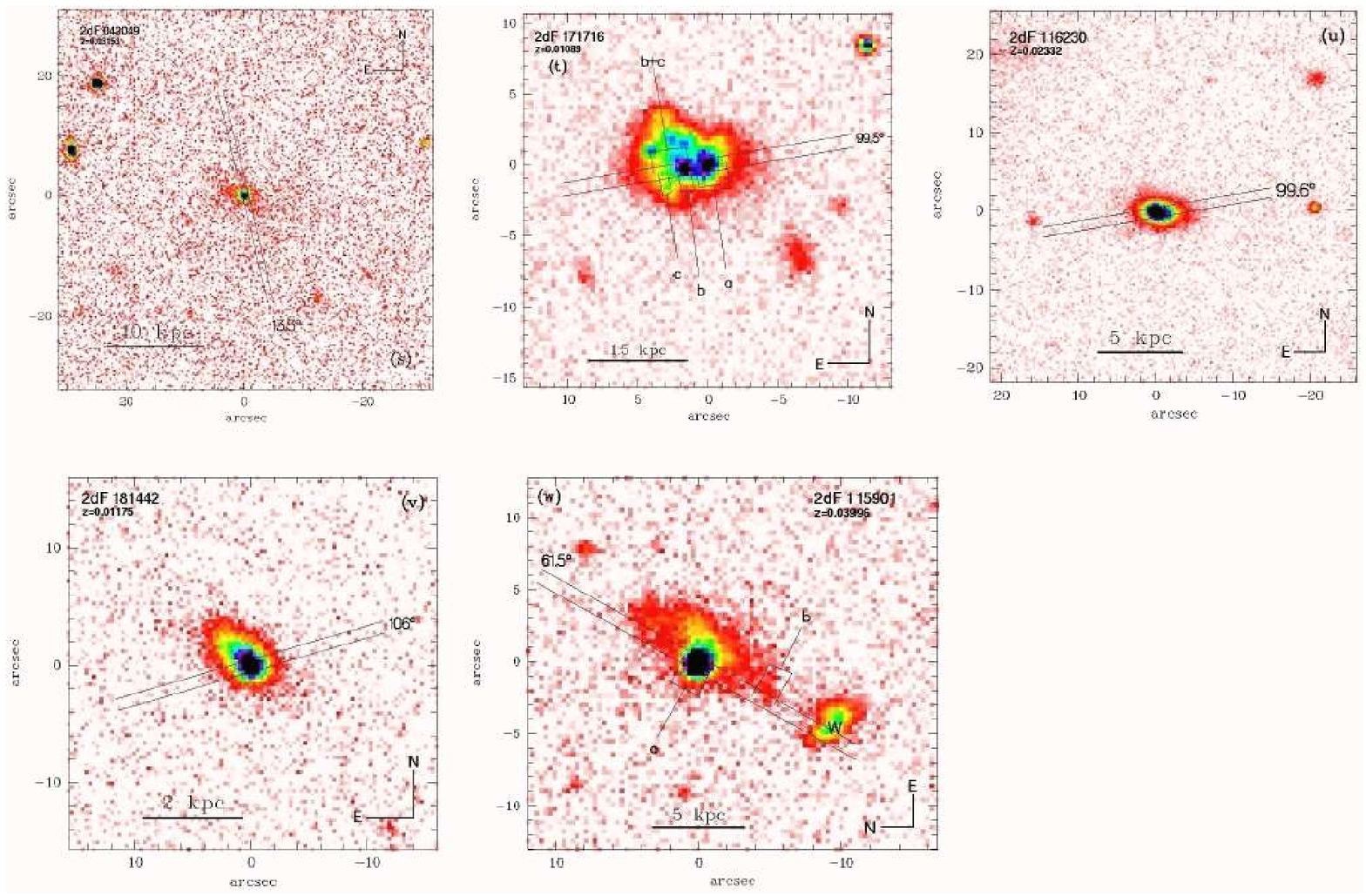,height=12.0cm,angle=0.,clip=}}}
\end{picture}
\caption{continued}
\end{figure*}

\subsection{Photometry \label{phot}}
%
All but one of the galaxies (\object{Tol 1400--411}) included in the spring 2003 sample were
observed in the $V$ band (5 min), whereas for the remaining galaxies both 
$V$ and $I$ exposures (5 min each) were taken. 
In either observing run we used EFOSC2 in its imaging mode 
and with a 2$\times$2 pixel binning, yielding an instrumental scale of 
0\farcs314 pixel$^{-1}$. 

The available exposures allow to derive surface brightness 
profiles (SBPs) down to typically $\mu= $26 \sbb\ in $V$, i.e. they are 
sufficiently deep for detecting the low-surface brightness (LSB) component 
in the sample galaxies. Standard imaging reduction procedures were carried out
using ESO-MIDAS\footnote{ESO-MIDAS is the acronym for the European Southern
Observatory Munich Image Data Analysis System which is developed
and maintained by the European Southern Observatory.}.

A reliable calibration using \citet{Lan92} standards could only
be established for two nights, April 24 and 25, 2003.
A rough calibration is probably also possible for the first half of the 
night on September 30, 2003.
For objects observed in the remaining nights the calibration had to rely by
necessity on data from the literature.
Part of these images was calibrated using total $B$ and $R$ magnitudes 
available from the SuperCOSMOS sky 
survey\footnote{available from {\tt http://www-wfau.roe.ac.uk/sss/}}.
The accuracy of these magnitudes may vary from case to case and can be
subject to systematic effects which are estimated to yield uncertainties 
of up to a few tenths of a magnitude. 
Additional uncertainties are introduced by the conversion of $B$ and $R$ magnitudes 
to $V$ magnitudes for which we assumed throughout ($B-V$) $\approx 0.6\times$ ($B-R$). 
This approximative relation was derived by fitting evolutionary synthesis models computed 
with the PEGASE 2.0 code \citep{FRV97} and refers to a stellar population with an 
age between 120 Myr and 10 Gyr forming with an exponentially decreasing 
star formation rate with an e-folding time of 1 Gyr.
In those models we furthermore assumed a Salpeter initial mass function
and a fixed stellar metallicity of $Z_{\odot}/5$. We did not take into account 
the effect of nebular line emission on broad band colours, as for the star formation
history and the age interval considered here it has a negligible effect on the integral colors.

Note that an integral ($B-V$)/($B-R$) colour ratio of $\approx$0.6 is also conform
with model predictions for a present-day Sd galaxy \citep[see, e.g., ][ for details]{Bicker2004}. 
An approximative ($B-V$)/($B-R$) ratio of $\approx$0.6 is also consistent with 
observations of emission-line galaxies \citep[see, e.g., ][ and references therein]{salzer89a} 
and approximates satisfactorily the integral colours of BCDs.
For instance, the mean $B-V$ and $V-R$ LSB colours inferred from the BCD sample 
of \citet{Cairos01b} are 0.45 mag and 0.33 mag, respectively, yielding 
a $B-V$/$B-R$ ratio of $\approx$0.6. 
The same colour ratio is obtained from the sample of Virgo dwarf ellipticals with 
available $BVR$ photometry, studied in \citet{vanZee2004}.

For \object{Tol 1334--326} and \object{Tol 1304--386} for which SuperCOSMOS data are
not available, the $V$ magnitude has been estimated from the $B$ and $R$ 
magnitudes given in \citet{GdP2003} and \citet{LV89}, respectively. 
Likewise, we have estimated the $V$ magnitude of \object{2dF 181442} from 
the $B$ magnitude of \citet{Maddox90}, assuming a mean $B-V$ of 0.4 mag. 
As no $I$ band magnitudes were available in the literature
for the latter system and for \object{2dF 116230}, we have vertically shifted their 
$I$ band SBPs such as to match the central intensity 
of $V$ band SBPs.
This allows us to quantify the colour variation between 
the star-forming component and the LSB host galaxy.
Finally, the $V$ band exposure of \object{Tol 1924--416} was calibrated using 
archival {\sl HST} WFPC2 data (Proposal ID 6708, PI: \"Ostlin) 
in $V$ and $I$ (filters F555W and F814W, respectively).

Surface photometry has been carried out for all sample galaxies except for
\object{2dF 042049} and \object{Tol 1400--411}. 
The first system is too compact for its LSB host to be studied 
quantitatively and the second one could not be observed due to scheduling constraints.
For a detailed discussion of the latter system the reader is referred, however, 
to \citet{Kai2003} who investigated its photometric structure 
using deep near infrared $JHK\arcmin$ data. 
Surface brightness profiles (Fig. ~\ref{SBPs}) were derived 
using method {\tt iv} described in \citet{Papaderos2002} and
decomposed into the star-forming and LSB components
following the procedure detailed in \citet{Kai2003}. 

Table \ref{tab:phot} lists photometric quantities for 21 galaxies in
our sample. All tabulated values are corrected for Galactic absorption 
based on the extinction maps by \citet{Schlegel89} available in 
NED\footnote{NASA/IPAC Extragalactic Database;\\ {\tt http://nedwww.ipac.caltech.edu}}.
In the first column of Table \ref{tab:phot} we list the name and morphological type
of each target according to the BCD classification scheme of \citet{LT86}.
Column 2 gives the extrapolated central surface brightness $\mu_{\rm E,0}$ (\sbb) 
and exponential scale length $\alpha$ (kpc) of the LSB host galaxy, obtained by 
fitting an exponential model to the outer part of each SBP. 
In col. 3 we tabulate the apparent m$_{\rm LSB}$ and absolute M$_{\rm LSB}$ magnitude 
of the LSB component in $V$, as derived from the exponential fits.
The apparent and absolute $V$-band magnitude, 
computed from SBP integration
are given in col. 4, and col. 5 lists the effective radius r$_{\rm eff}$ (kpc) and 
mean surface brightness $\mu_{\rm eff}$ inside r$_{\rm eff}$. 
In col. 6 we include the Petrosian radius r$_{\rm Petr}$ (kpc) defined as 
the radius where the Petrosian
$\eta$ function \citep{Petrosian76} decreases to a value of 1/3 
\citep[see, e.g., ][]{Takamiya99} and the apparent magnitude m$_{\rm Petr}$ 
within that radius.
Column 7 (first row) lists a light concentration index based on the 
radii r$_{80}$ and r$_{20}$ enclosing, respectively, 80\% and 20\% of the 
total $V$ emission. 
In addition, in col. 7 we tabulate the shape parameter $1/\eta$ of the S\'ersic model
which approximates best the observed $V$ SBP. 
The ratio r$_{25}$/r$_{\rm eff}$ of the isophotal 
radius at 25 $V$ \sbb\ and the effective radius are listed in col. 8. 
In the same column we indicate the radius interval in which exponential fits 
to the LSB emission were applied.
The fractional contribution of the LSB component to the total 
$V$ light and, when available, the mean $V-I$ colour within the fitted 
radius range are tabulated in col. 9.
In the last column of Table \ref{tab:phot} we include
the adopted distance in Mpc and the $V$ band Galactic
absorption in mag. 
Note that distances have been derived after correction of the measured redshifts 
(Table \ref{obs}) for the motion relative to the Local Group centroid and 
the Virgocentric flow \citep{K86}, assuming a Hubble constant 
of 75 km s$^{-1}$ Mpc$^{-1}$.

\section{Spectroscopic properties \label{res}}
%
\setcounter{figure}{1}
\begin{figure*}[hbtp]
   \hspace*{-0.2cm}\psfig{figure=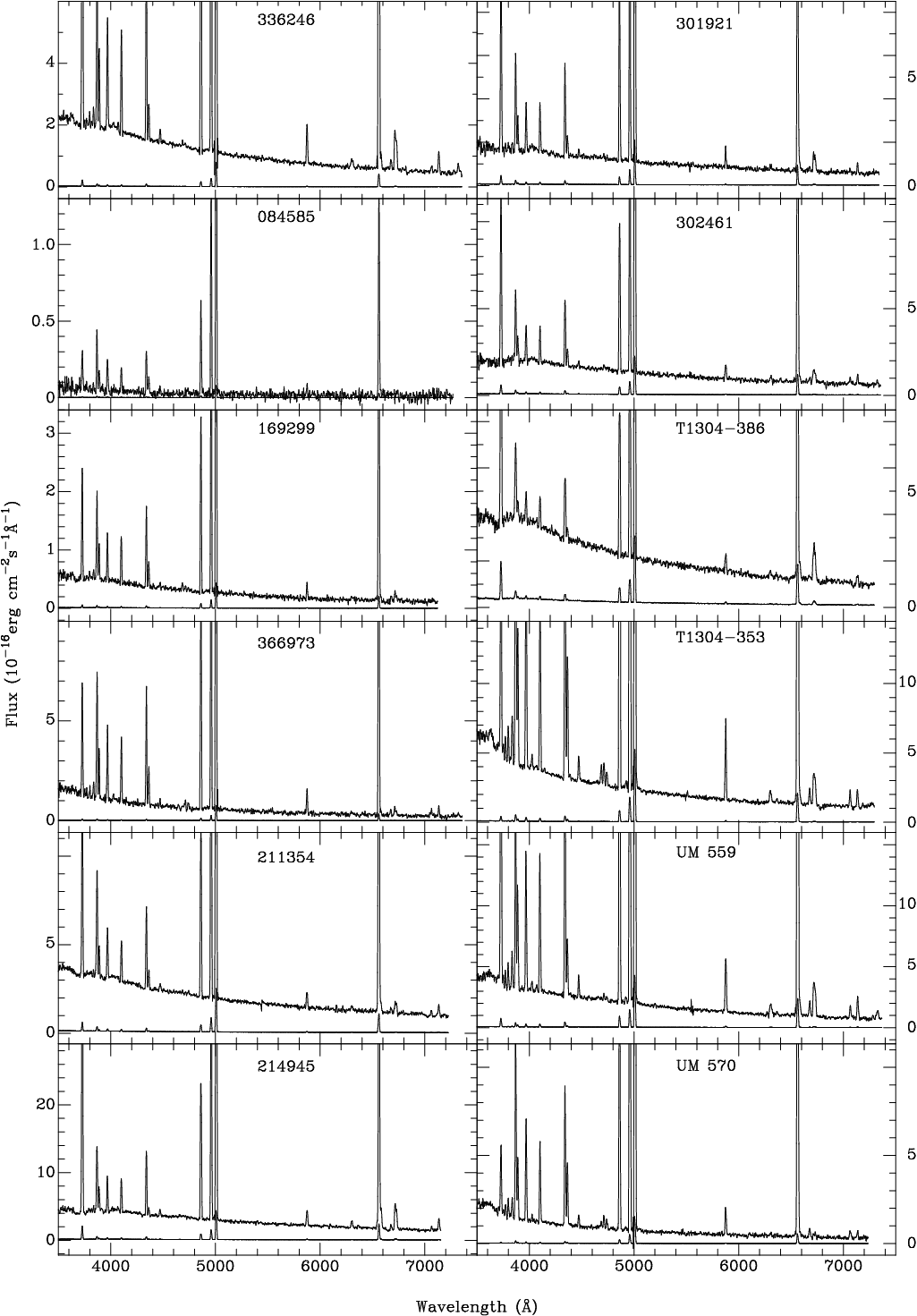,angle=0,width=17.0cm,clip=}
    \caption{Spectra of the brightest H {\sc ii} region in each sample galaxy
(labeled {\tt a} in Fig. \ref{Slit}), reduced to zero redshift. 
In each panel we include the observed spectrum downscaled by different factors
for a better visualization of strong emission lines.
      }
    \label{spectra1}
\end{figure*}

\setcounter{figure}{1}
\begin{figure*}[hbtp]
   \hspace*{-0.2cm}\psfig{figure=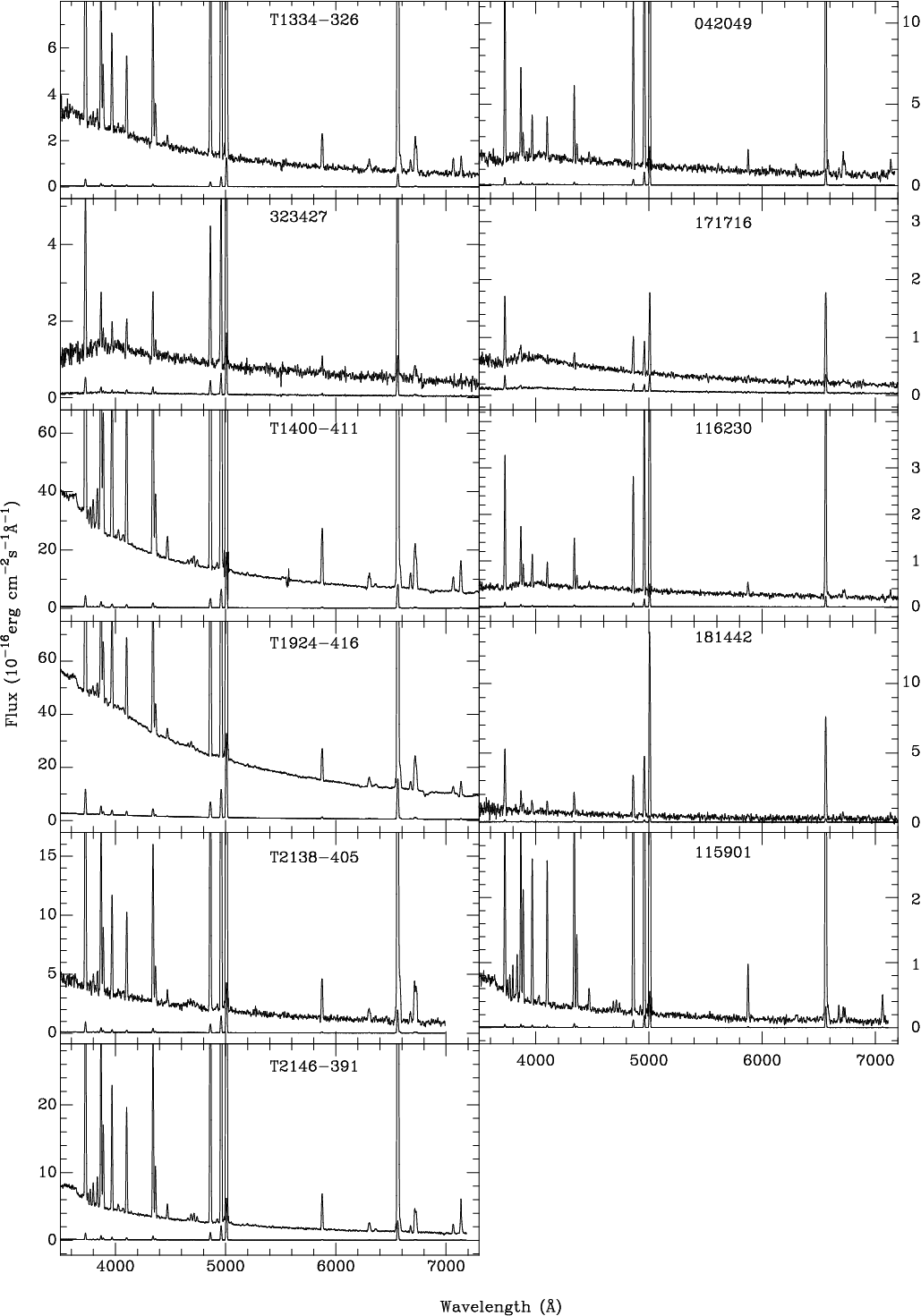,angle=0,width=17.0cm,clip=}
    \caption{continued.}
\end{figure*}

\subsection{Chemical abundances \label{chem}}
The spectra of the brightest star-forming regions in our sample  galaxies
are characterized by strong nebular emission lines which are superposed on a 
blue stellar continuum (Fig.~\ref{spectra1}). 
Twelve out of the 23 studied systems
(those classified nE BCD) possess a single dominant H\,{\sc ii} region, while 
in the remaining ones at least a second H\,{\sc ii} region can be distinguished.
In Fig.~\ref{spectra1} we include only the spectrum of the brightest H\,{\sc ii} 
region in each galaxy (region labeled {\tt a} in Fig. \ref{Slit}).

Emission line fluxes were measured using Gaussian profile fitting. 
The errors of the line fluxes were calculated from the photon
statistics in the non-flux-calibrated spectra. These uncertainties
have been propagated in the calculations of the elemental abundance errors.
Redshifts were derived using the observed wavelength of
$\sim$10 to $\sim$20 strong emission lines in each spectrum.
The redshift errors (Table \ref{obs}) are calculated as the square root mean
deviation of determinations from individual lines from the mean redshift.
The observed emission line fluxes $F$($\lambda$) relative to the H$\beta$  
fluxes, equivalent widths {\bb ($EW$s)} of emission lines, extinction coefficients $C$(H$\beta$),
observed H$\beta$ fluxes, and {\bb $EW$s} of the hydrogen absorption lines 
are listed in Table \ref{intensity} for the brightest H\,{\sc ii} region of each galaxy.

The electron temperature $T_{\rm e}$, ionic and total heavy element abundances 
were derived following \citet{ITL94,ITL97} and \citet{til95}. 
Because of the low spectral resolution of our data the [S {\sc ii}]$\lambda$6717 
and $\lambda$6731 lines are blended. Therefore, we adopted throughout for 
the electron number density $N_{\rm e}$(S {\sc ii}) a value of 100 cm$^{-3}$.
Note that the precise value of the electron number density 
makes little difference in the derived abundances
since in the low-density limit which holds for the H {\sc ii} regions
considered here, the element abundances do not depend sensitively 
on $N_e$.
The electron temperatures $T_{\rm e}$(O {\sc iii}) and $T_{\rm e}$(O {\sc ii}) 
for the high- and low-ionization zones in H {\sc ii} regions, respectively, 
ionization correction factors (ICF), ionic and total heavy element abundances 
for oxygen, neon, argon and, when possible, for iron, 
in the brightest H {\sc ii} regions (labeled ``a'' in 
Fig. \ref{Slit}) are included in Table \ref{heavy}.
Because the [N {\sc ii}] $\lambda$6584 line is blended 
with H$\alpha$ the nitrogen abundance could not be determined
with sufficient accuracy using a two-gaussian deblending procedure. 
Therefore, we prefer not to include it in Table \ref{heavy}.
Table \ref{t:HIIadd} contains a synoptic list of the spectroscopic properties of each 
H {\sc ii} region selected along the slit in Fig. \ref{Slit}.

\begin{figure}
\hspace*{0.0cm}\psfig{figure=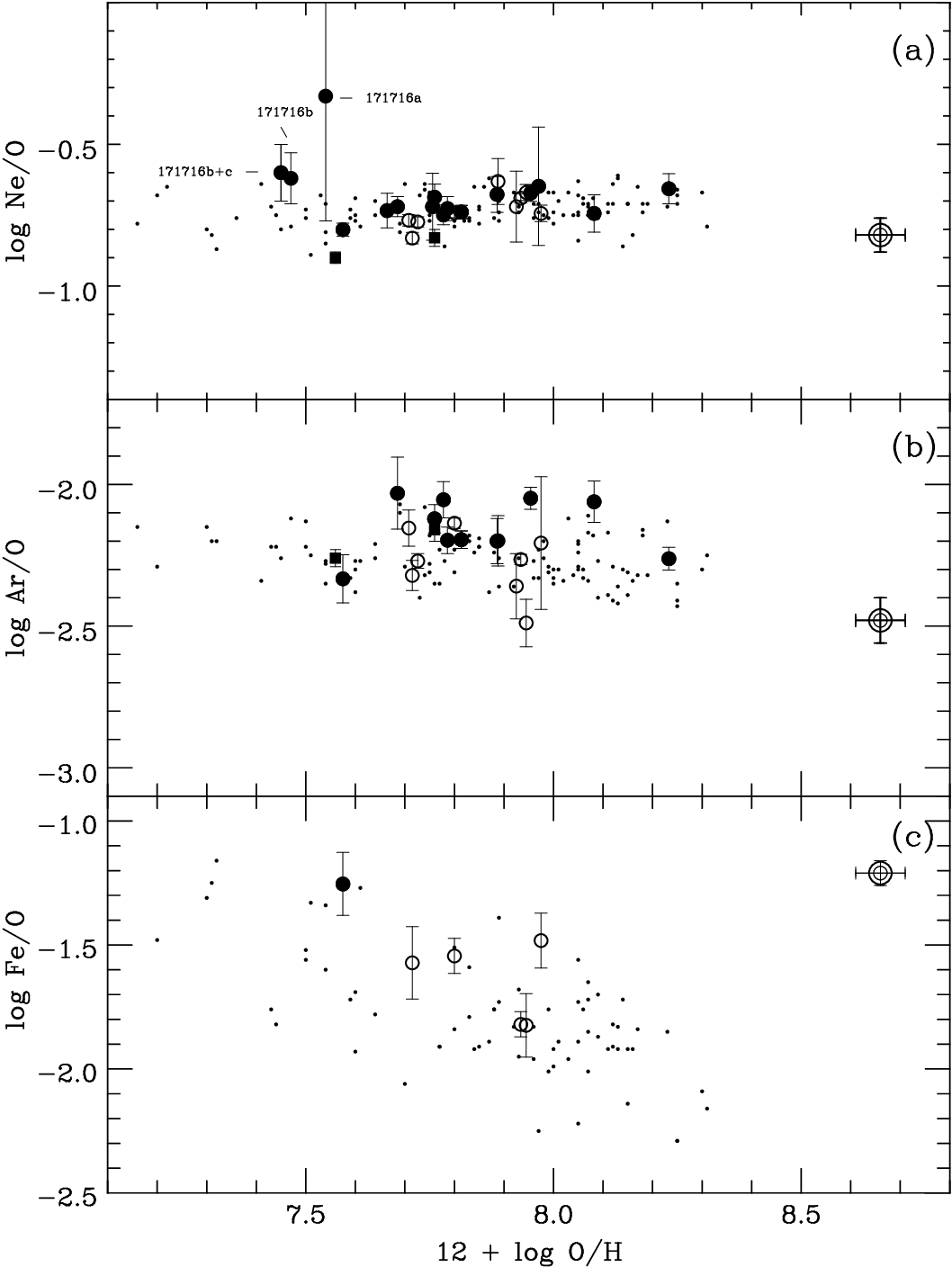,angle=0,width=8.5cm}
\caption{log Ne/O {\bf (a)}, log Ar/O {\bf (b)} and log Fe/O {\bf (c)} 
vs oxygen abundance 12 + log O/H for two samples of 
emission-line galaxies.
Large filled circles show 2dF galaxies, open circles Tololo and UM 
galaxies from our sample, and dots galaxies from the comparison sample. 
Data for \object{Tol 1214-277} \citep{Iz04} and 
Pox 186 \citep{Guseva04} are shown  by filled squares. 
Three H {\sc ii} regions in the galaxy 2dF~171716 are labeled.
Solar ratios from \citet{Asplund04} are indicated by double large open circles.
}
\label{nearfe}
\end{figure}

In Fig.~\ref{nearfe} we plot the log(Ne/O), log(Ar/O) and log(Fe/O)
abundance ratios vs. oxygen abundance for the present and
comparison samples. 2dFGRS galaxies are shown by large filled circles 
and Tololo and UM galaxies by open circles. 
We also include as filled squares \object{Tol 1214-277} and \object{Pox 186}, both 
observed in the present runs but discussed in the separate papers 
by \citet{Iz04} and \citet{Guseva04}, respectively.

For comparison, we show by dots in Fig.~\ref{nearfe} a sample of $\sim$100 
well-studied low-metallicity BCDs from \citet{ISGT04}. 
All spectra in the present and comparison sample were reduced
in a homogeneous manner according to the \citet{ITL94,ITL97} prescriptions. 
We adopted the solar abundances from \citet{Asplund04} 
(double large open circles in Fig.~\ref{nearfe}.
The best determined ratio, Ne/O, increases slightly with increasing O/H 
(by $\sim$0.1 dex in the considered metallicity range) which, 
following \citet{I06}, could be explained by a stronger depletion 
of oxygen onto dust grains in higher-metallicity galaxies.
At the same time, the Fe/O abundance ratio shows an underabundance 
of iron relative to oxygen as compared to solar, suggesting 
depletion of iron onto dust grains which is especially 
large for the high-metallicity galaxies.
On the other hand, no evident depletion effects are present in the Ar/O 
abundance ratio vs O abundance diagram. This is because a) the depletion
of oxygen is small b) Ar is noble gas and is not locked in grains and
c) the scatter of the Ar/O ratio is much larger than that of the Ne/O ratio 
and masks a possible weak trend with the oxygen abundance.

\begin{figure}
   \hspace*{0.7cm}\psfig{figure=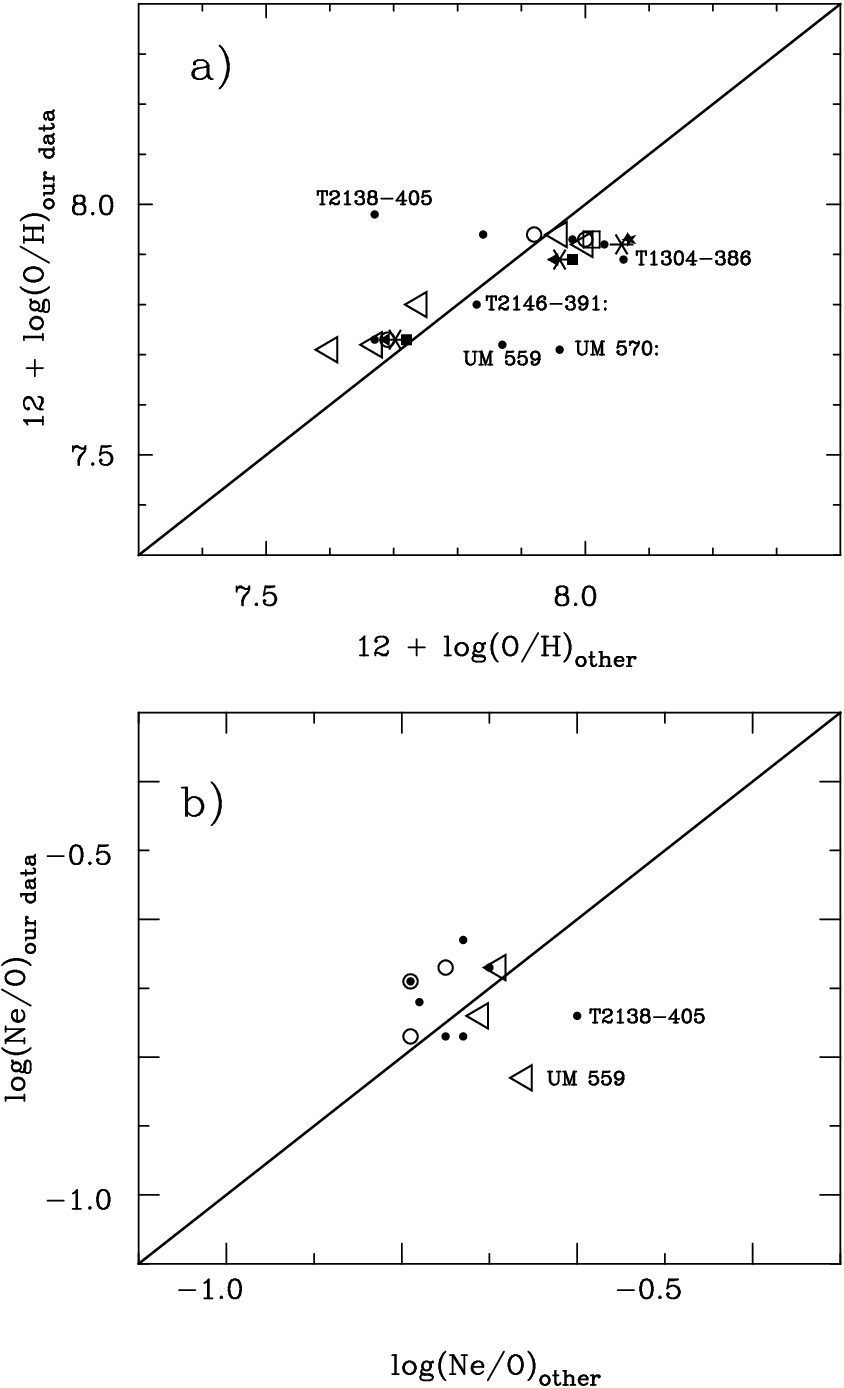,angle=0,width=7.cm}
    \caption{({\bf a}) Comparison of our derived oxygen abundances for 
    Tololo and UM galaxies with those of  
    \citet{terl91} ({\it filled circles}), \citet{Masegosa94}
    ({\it open circles}), \citet{kehrig04} and 
    \citet{telles95}  
    ({\it open triangles}), \citet{kobul96}  
    ({\it filled squares}), \citet{Pagel86} ({\it filled triangles}),
    \citet{Campbell92} ({\it asterisks}),
    \citet{Hid03} ({\it filled star}), 
    \citet{richer95} ({\it open square}).
    ({\bf b}) The same as in {\bf a} but for neon abundances.
    In both panels, equality between the literature data and ours is 
indicated by the diagonal line. Galaxies with the largest 
discrepancies and uncertainties are labeled.
      }
    \label{compare}
\end{figure}

Overall, we see in Fig.~\ref{nearfe} a good overlap  
of our sample galaxies with the comparison sample.
An exception is \object{2dF 171716} (labeled 171716a) which {\bb shows} a conspicuously 
large log(Ne/O) ratio, presumably because of a large error caused  
by the low signal/noise ratio of the available spectrum.
For this galaxy we extracted in addition 1D spectra of three regions, 
marked in Fig.~\ref{Slit} as {\tt b}, {\tt c} and {\tt b+c}. 
Note that the weak neon line could not be measured in region {\tt c}. 


\begin{table*}[tbh]
\caption{Spectroscopic properties of H\,{\sc ii} regions selected along the slit.}
\label{t:HIIadd}
\begin{tabular}{lcccrcc} \hline \hline
H\,{\sc ii} Region &$t_e$(O$^{++}$)$^{\rm a}$&12 + log(O/H)&$EW$(H$\alpha$)~(\AA)& 
F(H$\beta$)$^{\rm b}$&$C$(H$\beta$) dex &
100$\times$$F$([O {\sc iii}]4363)/$F$(H$\beta$) \\ \hline

336246{\bf a}           &1.53 $\pm$0.02 &7.81 $\pm$ 0.01&  754 $\pm$2  &138.6 $\pm$0.5 &0.060 $\pm$ 0.019 & 9.9 $\pm$ 0.3 \\
084585{\bf a}           &1.90 $\pm$0.08 &7.66 $\pm$ 0.04& 1531 $\pm$18 &7.2   $\pm$0.1 &0.000 $\pm$ 0.034 &20.4 $\pm$ 1.5 \\
084585{\bf b}           &1.75 $\pm$0.13 &7.73 $\pm$ 0.07&  893 $\pm$22 &6.1   $\pm$0.2 &0.505 $\pm$ 0.056 & 13.4 $\pm$ 1.9 \\
084585{\bf c}$^{\rm c}$&1.62 $\pm$0.03 &7.72 $\pm$ 0.03&  229 $\pm$10  &2.3   $\pm$0.2 &0.000 $\pm$ 0.102 &        ...    \\
169299{\bf a}           &1.75 $\pm$0.04 &7.68 $\pm$ 0.02&  737 $\pm$3  &35.8  $\pm$0.3 &0.000 $\pm$ 0.021 &14.5 $\pm$ 0.6 \\
169299{\bf b}$^{\rm c}$ &1.54 $\pm$0.10 &7.70 $\pm$ 0.09&  276 $\pm$15 &0.5   $\pm$0.1 &0.000 $\pm$ 0.401 &        ...    \\
366973{\bf a}           &1.66 $\pm$0.04 &7.78 $\pm$ 0.02& 1326 $\pm$5  &92.0  $\pm$0.8 &0.165 $\pm$ 0.022 &14.4 $\pm$ 0.6 \\
211354{\bf a}           &1.41 $\pm$0.02 &7.95 $\pm$ 0.02&  310 $\pm$1  &94.1  $\pm$0.5 &0.145 $\pm$ 0.020 &10.0 $\pm$ 0.4 \\
214945{\bf a}           &1.11 $\pm$0.03 &8.23 $\pm$ 0.03&  474 $\pm$1  &137.3 $\pm$0.7 &0.310 $\pm$ 0.020 & 4.3 $\pm$ 0.3 \\
301921{\bf a}           &1.69 $\pm$0.06 &7.76 $\pm$ 0.03&  410 $\pm$2  &53.7  $\pm$0.4 &0.265 $\pm$ 0.021 &13.1 $\pm$ 0.8 \\
302461{\bf a}           &1.55 $\pm$0.04 &7.79 $\pm$ 0.02&  438 $\pm$1  &66.7  $\pm$0.5 &0.250 $\pm$ 0.021 &10.1 $\pm$ 0.5 \\
T1304--386{\bf a}        &1.46 $\pm$0.08 &7.89 $\pm$ 0.05&  250 $\pm$1 &55.5  $\pm$0.5 &0.265 $\pm$ 0.022 & 7.9 $\pm$ 0.9 \\
T1304--386{\bf b}        &1.46 $\pm$0.05 &7.88 $\pm$ 0.03&  639 $\pm$3 &36.6  $\pm$0.3 &0.360 $\pm$ 0.022 & 8.1 $\pm$ 0.6 \\
T1304--353{\bf a}        &1.79 $\pm$0.02 &7.73 $\pm$ 0.01& 1424 $\pm$2 &395.6 $\pm$0.9 &0.240 $\pm$ 0.019 &17.3 $\pm$ 0.3 \\
UM 559{\bf a}           &1.58 $\pm$0.02 &7.72 $\pm$ 0.01& 1702 $\pm$2  &357.2 $\pm$0.9 &0.240 $\pm$ 0.019 & 9.6 $\pm$ 0.2 \\ 
UM 570{\bf a}           &1.83 $\pm$0.02 &7.71 $\pm$ 0.01& 1543 $\pm$4  &172.7 $\pm$0.6 &0.115 $\pm$ 0.019 &20.2 $\pm$ 0.4 \\
T1334--326{\bf a}        &1.49 $\pm$0.12 &7.92 $\pm$ 0.08&  798 $\pm$6 &105.5 $\pm$2.1 &0.265 $\pm$ 0.033 &11.2 $\pm$ 0.2 \\
T1334--326{\bf b}        &1.52 $\pm$0.03 &7.92 $\pm$ 0.02& 1394 $\pm$4 &83.7  $\pm$0.5 &0.325 $\pm$ 0.020 & 11.6 $\pm$ 0.4 \\
T1334--326{\bf c}        &1.56 $\pm$0.03 &7.83 $\pm$ 0.02& 1227 $\pm$4 &73.5  $\pm$0.5 &0.085 $\pm$ 0.021 & 11.1 $\pm$ 0.5 \\
323427{\bf a}           &1.47 $\pm$0.11 &7.76 $\pm$ 0.07&  277 $\pm$2  &31.6  $\pm$0.4 &0.180 $\pm$ 0.027 & 7.3 $\pm$ 1.2 \\
T1400--411{\bf a}        &1.45 $\pm$0.01 &7.93 $\pm$ 0.01& 1138 $\pm$1 &1705.0$\pm$2.2 &0.230 $\pm$ 0.018 &10.2 $\pm$ 0.2 \\
T1400--411{\bf b}        &1.41 $\pm$0.01 &7.95 $\pm$ 0.01& 1080 $\pm$1 &779.7 $\pm$1.6 &0.205 $\pm$ 0.019 & 8.7 $\pm$ 0.2 \\
T1400--411{\bf c}        &1.10 $\pm$0.13 &8.09 $\pm$ 0.12& 1094 $\pm$7 &33.1  $\pm$0.6 &0.070 $\pm$ 0.030 & 3.8 $\pm$ 0.9 \\
T1924--416{\bf a}        &1.42 $\pm$0.02 &7.94 $\pm$ 0.02&  419 $\pm$1 &973.1 $\pm$3.3 &0.260 $\pm$ 0.019 & 8.7 $\pm$ 0.3 \\
T1924--416{\bf b}        &1.39 $\pm$0.01 &7.99 $\pm$ 0.01&  655 $\pm$1 &3273.0 $\pm$4.0 &0.290 $\pm$ 0.018 & 8.7 $\pm$ 0.2 \\
T2138--405{\bf a}        &1.38 $\pm$0.02 &7.98 $\pm$ 0.02& 1192 $\pm$2 &269.5 $\pm$1.2 &0.185 $\pm$ 0.019 & 8.9 $\pm$ 0.3 \\
T2146--391{\bf a}        &1.61 $\pm$0.02 &7.80 $\pm$ 0.01& 1547 $\pm$2 &460.1 $\pm$0.9 &0.195 $\pm$ 0.019 &12.4 $\pm$ 0.2 \\
042049{\bf a}           &1.28 $\pm$0.05 &8.08 $\pm$ 0.04&  507 $\pm$3  &82.5  $\pm$1.0 &0.165 $\pm$ 0.025 & 8.4 $\pm$ 0.8 \\ 
171716{\bf a}           &1.69 $\pm$0.57 &7.54 $\pm$ 0.26&  108 $\pm$2  &6.6   $\pm$0.2 &0.080 $\pm$ 0.045 & 5.8 $\pm$ 3.5 \\
171716{\bf b}           &1.80 $\pm$0.01 &7.47 $\pm$ 0.05&  349 $\pm$3  &8.1   $\pm$0.1 &0.095 $\pm$ 0.030 & 8.1 $\pm$ 0.8 \\   
171716{\bf c}           &2.00 $\pm$0.57 &7.38 $\pm$ 0.20&  216 $\pm$2  &4.0   $\pm$0.2 &0.175 $\pm$ 0.182 & 18.1 $\pm$ 7.5 \\  
171716{\bf b+c}         &1.78 $\pm$0.13 &7.45 $\pm$ 0.06&  280 $\pm$3  &16.0  $\pm$0.3 &0.060 $\pm$ 0.030 & 7.2 $\pm$ 0.9 \\  
116230{\bf a}           &1.47 $\pm$0.06 &7.89 $\pm$ 0.04&  354 $\pm$2  &23.1  $\pm$0.2 &0.080 $\pm$ 0.024 &10.6 $\pm$ 0.9 \\
181442{\bf a}           &1.32 $\pm$0.02 &7.97 $\pm$ 0.12&  381 $\pm$6  &24.7  $\pm$0.7 &0.165 $\pm$ 0.046 & 6.4 $\pm$ 2.0 \\
115901{\bf a}           &1.80 $\pm$0.03 &7.57 $\pm$ 0.02& 2255 $\pm$6  &70.2  $\pm$0.3 &0.125 $\pm$ 0.020 &12.7 $\pm$ 0.3 \\
115901{\bf b}           &     ...       &      ...      & 2250 $\pm$950&0.8   $\pm$0.4 &       ...        &        ...    \\   

  \hline

  \hline\hline
\end{tabular}
$^{\rm a}$$t_e$=10$^{-4}$$T_e$. \\
$^{\rm b}$observed flux in units of 10$^{-16}$ erg s$^{-1}$ cm$^{-2}$. \\
$^{\rm c}$electron temperature and oxygen abundance obtained with the empirical calibration 
by \citet{Pil2000}. \\
\end{table*}

Next, we compare our results for the Tololo and UM objects
with data from the literature.
Using our technique for abundance determination and data for line 
intensities and equivalent widths from
\citet{kehrig04}, \cite{terl91} and \citet{telles95} we recalculated
the ionic and element abundances for the Tololo and UM subsample.
For some objects, the data from previous studies is not complete.
For instance, the [O {\sc ii}] $\lambda$3727 intensity is not
available for UM\ 570 \citep{kehrig04,telles95} 
while the H$\alpha$ line intensity is not available for UM\ 570 and 
\object{Tol 2146--391} \citep{terl91}. 
In those cases, we have complemented the literature data by our own, 
after scaling the measured [O {\sc ii}]$\lambda$3727 and 
H$\alpha$ line intensities relative to H$\beta$.
In Fig.~\ref{compare} we compare the oxygen ({\bf a}) and neon
({\bf b}) abundances for Tololo and UM galaxies  with 
data by \citet{terl91} ({\it filled small circles}),
\citet{Masegosa94} ({\it open circles}) and 
\citet {kehrig04} and \cite{telles95}  
({\it open triangles}).
In addition, we compare our oxygen abundance determination 
for \object{Tol 1304--353} and \object{Tol 1304--386} with that of \citet{kobul96}
({\it filled squares}), \citet{Pagel86} ({\it filled triangles})
and \citet{Campbell92} and \citet{Marconi94} ({\it asterisks}),
for \object{Tol 1334--326} with \citet{Campbell92} ({\it asterisk}), and for
\object{Tol 1400--411} with \citet{Hid03} ({\it filled star}) 
and \citet{richer95} ({\it open square}).
A significant dispersion of the {\it filled small circles} 
about the  equality line in Fig.~\ref{compare} indicates that there exist
some discrepancies between our abundances and those calculated from the 
data of \citet{terl91}. 
The largest differences ($\sim$ 0.3 dex) are seen for \object{Tol 2138--405} and \object{UM 570}.
Otherwise our determinations are, within formal errors, in 
agreement with published values.

More specifically, we confirm the low metallicity (12 + log(O/H) $\sim$ 7.7) 
of \object{UM 559} and \object{UM 570} discussed by \citet{salzer89a,salzer89b}
and derived by \citet{kehrig04}. This is also the case for the 
oxygen abundance 12 + log(O/H) $\la$ 7.8 for \object{Tol 2146--391} inferred by 
\citet{kehrig04}.
With regard to \object{Tol 1304--353} we obtain a relatively low oxygen abundance, 12 + log(O/H) $\sim$ 7.7, 
in accordance with determinations by \citet{Masegosa94},
\citet{kobul96}, \citet{Pagel86} and \citet{Campbell92}.

In summary, our 2dFGRS galaxy sample contains two very metal-deficient 
systems with an oxygen abundance 12 +log(O/H) $\leq$ 7.6 and another 
7 galaxies with oxygen abundance 12 +log(O/H) $\la$ 7.8 (cf. Tables 
\ref{t:HIIadd} and \ref{heavy}).
We confirm previous oxygen abundance determinations for 
\object{Tol 1304--353}, \object{Tol 2146--391}, 
\object{UM 559} and \object{UM 570} to be 12 +log(O/H) $\leq$ 7.8.

\begin{figure}
\hspace*{0.0cm}\psfig{figure=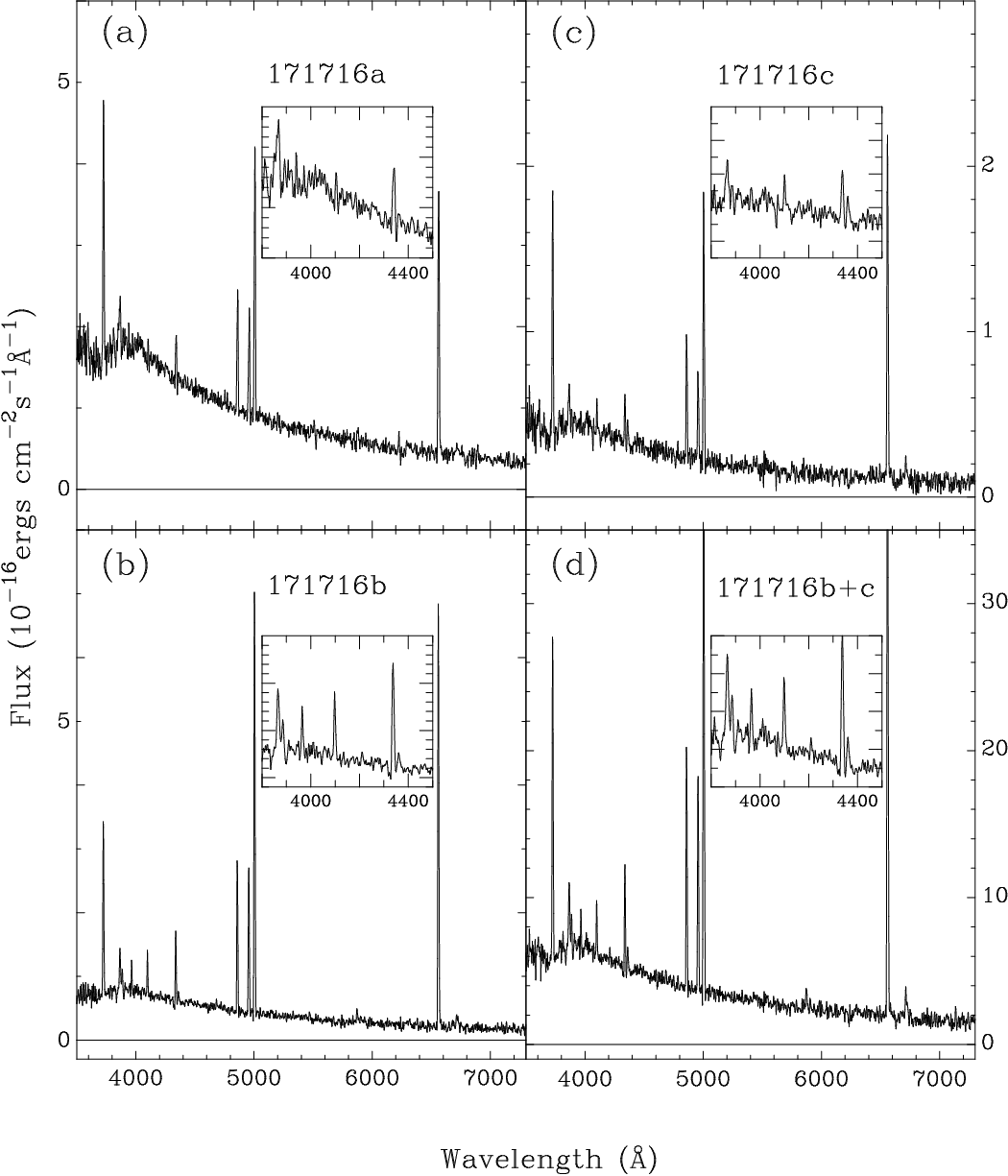,angle=0,width=9.cm}
    \caption{
Extinction-corrected restframe spectra of H\,{\sc ii} regions
in the metal-deficient galaxy 2dF~171716 (cf. Fig. \ref{Slit}t).
The insets display enlarged parts of spectra with weak lines, 
including the [O {\sc iii}]$\lambda$4363\AA\ emission line.
      }
    \label{171716}
\end{figure}

\subsection{New low-metallicity galaxies from the 2dFGRS \label{lowZ}}
The present spectroscopic study of the 2dF Galaxy Redshift Survey 
led to the discovery of two new extremely metal-deficient
emission-line galaxies, \object{2dF 171716} and \object{2dF 115901}, with an oxygen abundance of
12+log(O/H) $\sim$ 7.5 and $\sim$ 7.6, respectively.
Thus, the gap between the three most metal-deficient BCDs known, 
SBS 0335--052\,W\&E \citep[12+log(O/H)=7.12 and 7.3, ][]{ITG05,I97},
and I\,Zw\,18 \citep[12+log(O/H)=7.2, ][ and references therein]{SS1972,ICF99} 
and the main population of {\bb star-forming}
dwarf galaxies with a mean oxygen abundance 
12+log(O/H)$>$7.6 is gradually filling.

The irregular star-forming galaxy \object{2dF 171716} (Fig. \ref{Slit}t)
contains two bright H{\sc ii} regions located at the southeastern 
part of its blue, extended high-surface brightness component.
In Fig.~\ref{171716} we include the spectra of four emission-line regions 
along the slit labeled in Fig.~\ref{Slit}t {\tt a}, {\tt b}, {\tt c} and {\tt b+c}. 
The electron temperature, ionic and heavy element abundances 
of the brightest H\,{\sc ii} region {\tt  a} (labeled 
171716a in Fig.~\ref{nearfe}) are subject to large uncertainties 
(Table~\ref{t:HIIadd}). 
In the second brightest H\,{\sc ii} region {\tt b}, however, where 
the signal-to-noise is higher we were able to 
obtain a more reliable oxygen abundance determination. 
The value 12 + log(O/H) = 7.38 $\pm$ 0.20 
inferred for the extended and by $\sim$0.8 mag 
fainter region {\tt c} should be considered with caution, 
as this could be affected by the low S/N 
[O {\sc iii}]$\lambda$4363/H$\beta$ intensity ratio.
Relatively larger line fluxes and, correspondingly, 
smaller uncertainties in the heavy element abundances 
(Table~\ref{t:HIIadd}) are obtained, however, when 
a 1D-spectrum is extracted within a larger
1\arcsec $\times$ 2\farcs77 aperture, which includes both 
regions {\tt b} and {\tt c} (labeled b+c in Fig.~\ref{Slit}).
From the analysis of all 1D-spectra
in Fig.~\ref{171716}, we can constrain the mean oxygen abundance 
of \object{2dF 171716} to be 12+log(O/H) $\sim$ 7.5.

\begin{figure*}[!ht]
\begin{picture}(24,20)
\put(0.0,0.0){{\psfig{figure=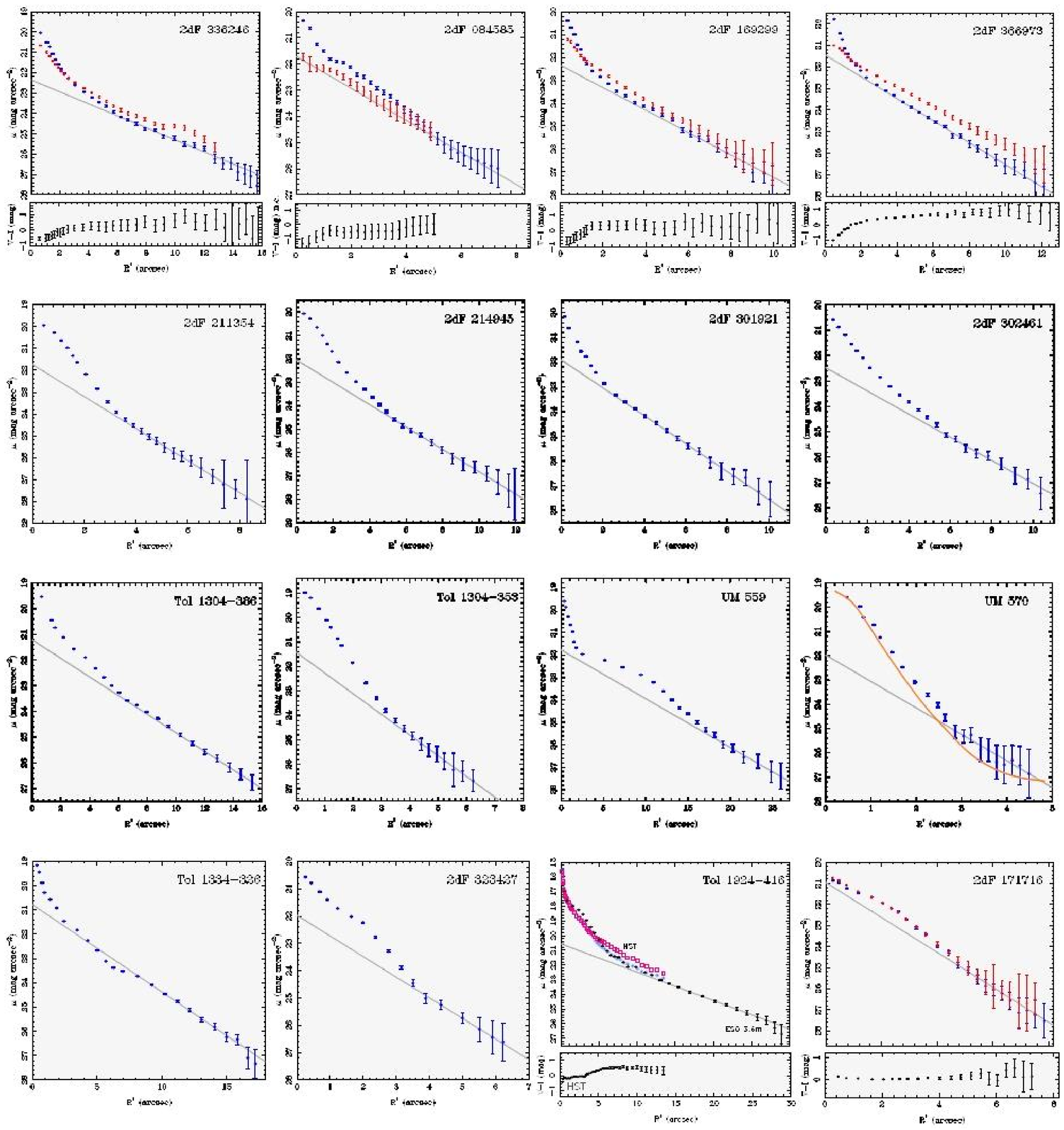,height=19.5cm,angle=0.0,clip=}}}
\end{picture}
\caption{$V$ band surface brightness profiles (SBPs; filled circles)
for all sample galaxies, except for\object{Tol 1400--411} and \object{2dF 042049}. 
Solid lines show exponential fits to the LSB component (Table \ref{tab:phot}). 
Whenever $I$ band data are available (open symbols) we also
include radial $V-I$ colour profiles beneath each SBP. 
For \object{Tol 1924--416}, for which only $V$ ground-based data are available, we 
include SBPs derived from archival {\sl HST} WFPC2 $V$ and $I$ images. 
The $I$ SBPs of \object{2dF 116230} and \object{2dF 181442} 
for which no calibration is available, have been shifted vertically such as to match the 
central intensity in $V$. {\bb The uncalibrated $V-I$ profiles of these systems
serve only to illustrate colour gradients.}}
\label{SBPs}
\end{figure*}

\setcounter{figure}{5}
\begin{figure*}[!ht]
\begin{picture}(24,10)
\put(0.0,0.0){{\psfig{figure=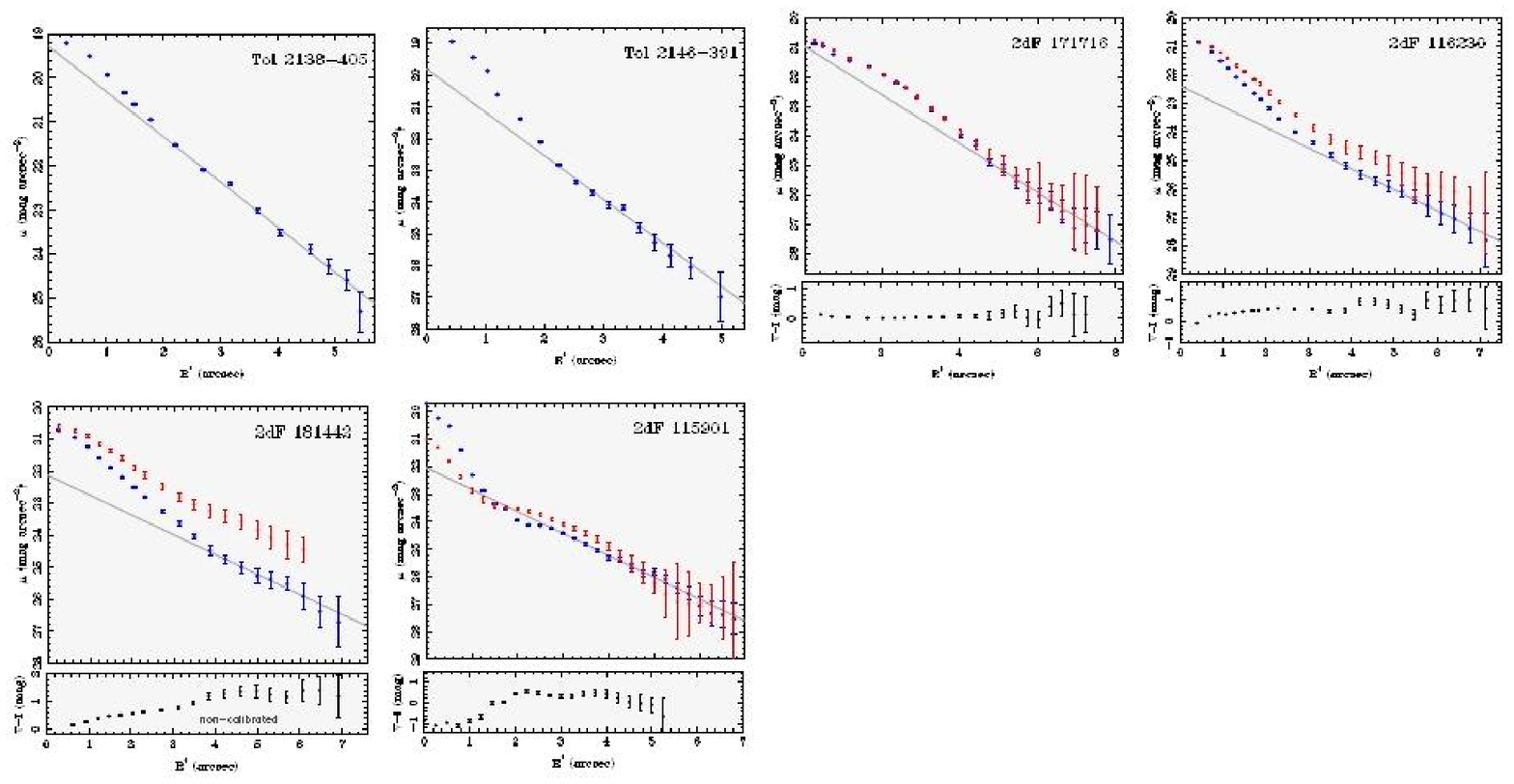,height=9.8cm,angle=0.0,clip=}}}
\end{picture}
\caption{continued.}
\end{figure*}

The BCD \object{2dF 115901} is the second most metal-poor galaxy in our 2dFGRS sample. 
Its bright (M$_V$=--13.9 mag), compact H\,{\sc ii} region 
{\tt a} (cf. Fig. \ref{Slit}w) with an oxygen abundance 
12 + log (O/H) = 7.57 $\pm$ 0.02 is the locus of copious
ionized gas emission. The very large equivalent width of 
the [O {\sc iii}]$\lambda$5007 emission line in this region, 
$EW$([O {\sc iii}]$\lambda$5007) = 2036 \AA\ (Table~\ref{intensity}),
suggests that more than 50\% of the $V$-band emission originates from 
this line. The extreme importance of ionized gas emission to the 
luminosity budget of region {\tt a} is also reflected on its 
measured EW(\ha), exceeding 2200 \AA.
In the adjacent region {\tt b}, the H$\beta$ flux is $\sim$ 90 times smaller
than in region {\tt a} and the [O {\sc iii}] $\lambda$4363 emission line 
could not be detected, while the $EW$(H$\alpha$) remains very large (Table~\ref{t:HIIadd}). 
In the absence of direct electron temperature diagnostics, the metallicity of region {\tt b}
cannot be constrained with sufficient precision.

\section{Photometric properties \label{photo}}
The galaxy sample investigated here comprises mainly systems of the nE 
and iI morphological BCD types in the \citet{LT86} classification 
scheme. With the exception of \object{Tol\ 2138--405} (M$_V$=--19.8) and 
\object{Tol 1924--416} (M$_V\approx$--19.3) all targets qualify by their 
absolute magnitude (M$_V>$--18 mag) as genuine dwarf galaxies.
%
%
\begin{figure}
\begin{picture}(9.2,10.6)
\put(0.,10.6){{\psfig{figure=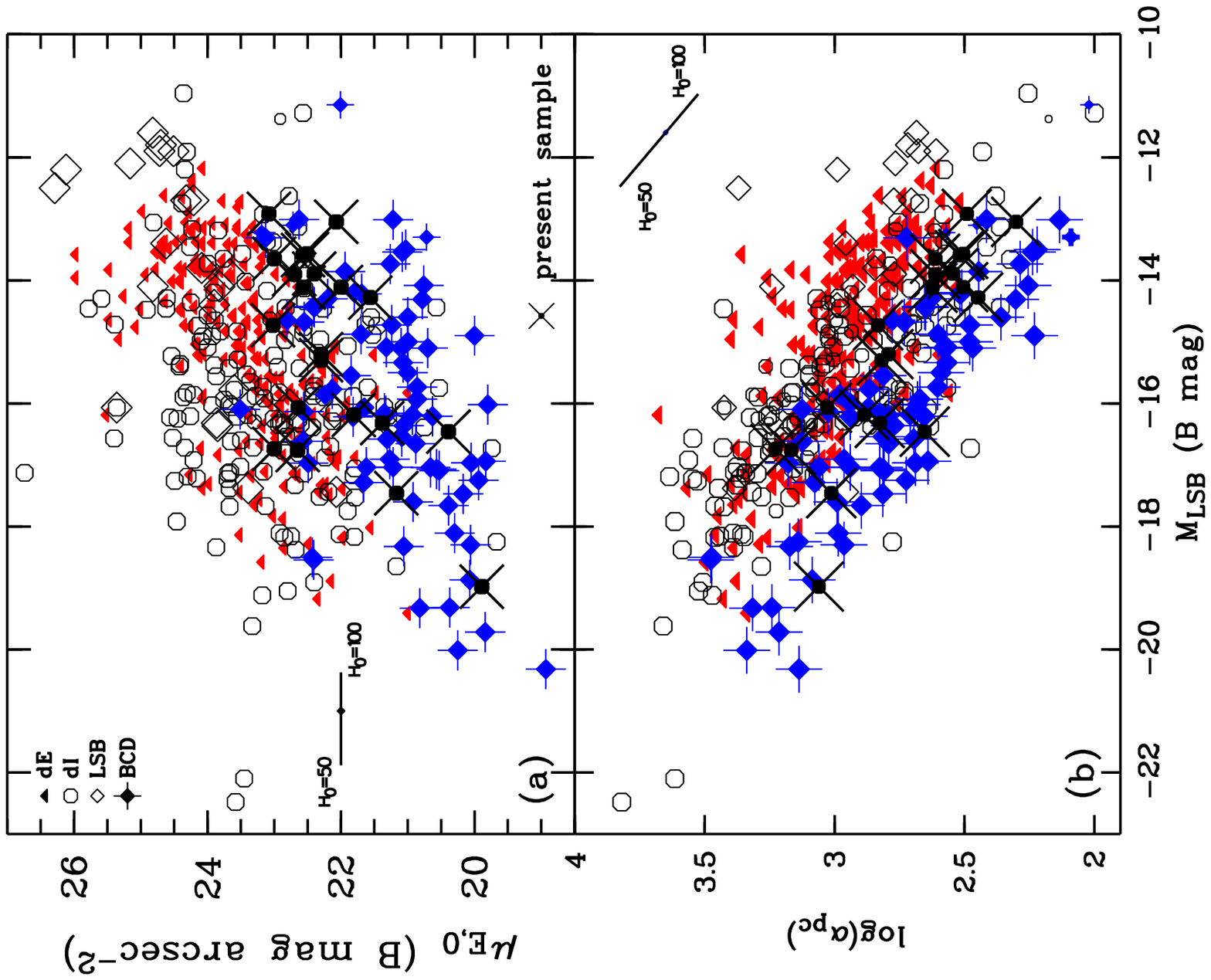,height=8.6cm,angle=-90.0,clip=}}}
\end{picture}
\caption{
Comparison of the structural properties of the LSB host 
of our sample galaxies with those for BCDs and other types dwarf galaxies.
Data for iE/nE BCDs are compiled from 
\citet{Cairos01a}, \citet{drinkwater91}, \citet{Mar97}, \citet{noeske99}, 
\citet{Kai00} and \citet{Papaderos96a}.
Data for other types of dwarf galaxies (dI: dwarf irregulars, dE: dwarf ellipticals, LSB: low-surface brightness)
are taken from \citet{binggeli91,binggeli93}, 
\citet{bothun91}, \cite{caldwell87}, \citet{carignan89}, \citet{Hopp91}, 
\citet{PT96}, \citet{vigroux86} and \citet{vZ00}.
Photometric quantities in $V$ have been converted into $B$ assuming 
a typical $B-V$ colour of 0.5 mag for the LSB component of our sample galaxies.
The lines show the shift of the data points caused by a change 
of the Hubble constant from 75 to 50 and 100 km s$^{-1}$ Mpc$^{-1}$.
{\bf (a)} Central surface brightness $\mu _{\rm E,0}$ vs. 
absolute $B$ magnitude $M_{\rm LSB}$ of the LSB component.
{\bf (b)} Logarithm of the exponential scale length $\alpha $ in pc 
vs. $M_{\rm LSB}$.}
\label{structure}
\end{figure}

\begin{figure}
\begin{picture}(8,6.8)
\put(-0.5,6.8){{
\psfig{figure=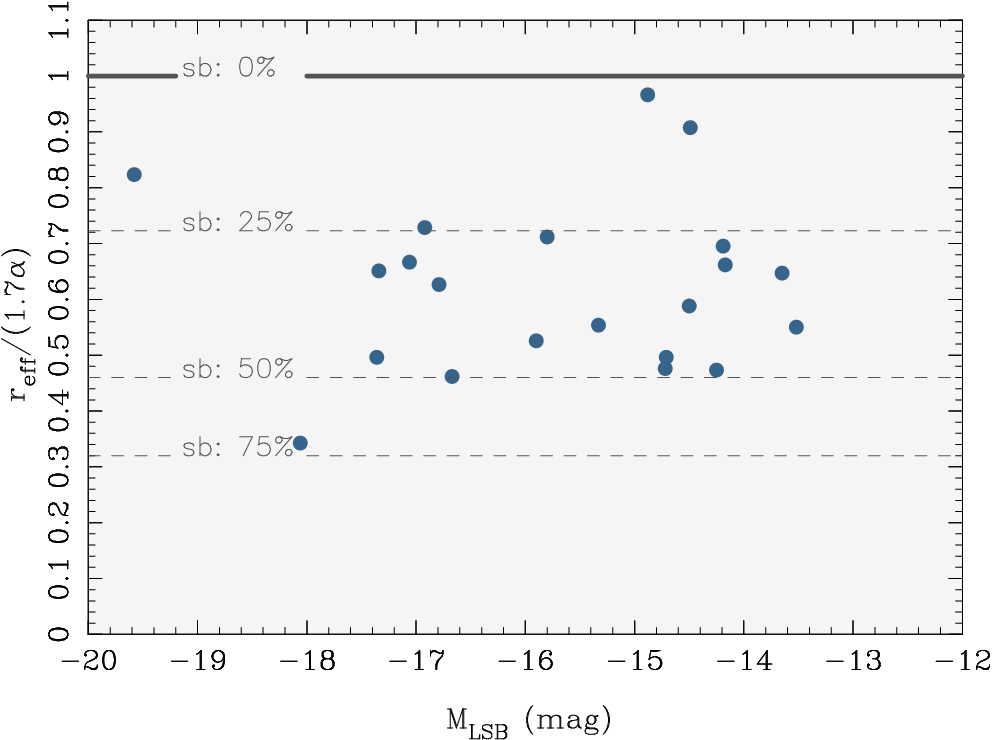,height=9.2cm,angle=-90.0,clip=}}}
\end{picture}
\caption{Absolute $V$ magnitude M$_{\rm LSB}$ of the host galaxy
vs. ratio r$_{\rm eff}$/1.7$\alpha$, where r$_{\rm eff}$ denotes 
the effective radius of our sample galaxies and $\alpha$ is the exponential 
scale length of the LSB host. A ratio r$_{\rm eff}$/1.7$\alpha\approx1$
is expected for a purely exponential intensity profile, that is, in the case 
when the luminosity contribution of the starburst is negligible.
Horizontal lines indicate the expected r$_{\rm eff}$/1.7$\alpha$ ratio 
when an exponentially distributed young stellar population with a scale length 
$\alpha$/4, {\bb providing 25\%, 50\% and 75\% of the total galaxy's light, 
is superposed on the LSB component.}}
\label{r_eff}
\end{figure}

This becomes clearer if instead of the total magnitude, one considers only 
the absolute magnitude M$_{\rm LSB}$ of the LSB component 
(col. 3 in Table \ref{tab:phot}), as derived by exponential 
fitting to the outer part of each SBP (cf. Fig. \ref{SBPs}). 
Excluding \object{Tol\ 2138--405}, we obtain in that case a mean 
M$_{\rm LSB}$ = --15.5$\pm$1.4 (with a standard deviation about the 
mean of $\sigma$=0.3), with \object{Tol 1942--416} (M$_{\rm LSB}\approx$--18 mag) being
the intrinsically brightest member in our sample. 

In most of our sample galaxies, the SBP of the outer parts of the 
LSB host is well approximated by an exponential law, as is generally true for dwarf galaxies. 
The exponential scale length $\alpha$ of the LSB component ranges between 
0.2 kpc (\object{2dF 366973}) and 1.7 kpc (\object{2dF 169299}) with a mean value 0.65$\pm$0.4 ($\sigma$=0.09) kpc. 
Thus, our sample contains no ultracompact systems with $\alpha\la$120 pc, as for 
example I\ Zw\ 18 \citep{Papaderos2002} and 
Pox 186 \citep[$\alpha=$120 pc, ][]{Guseva04} 
but rather displays structural properties typical for BCDs.
This is illustrated in Fig. \ref{structure} where the present galaxy sample 
is compared with other dwarf galaxies, such as dwarf irregulars (dIs) and 
dwarf ellipticals (dEs), in the M$_{\rm LSB}$ vs. log($\alpha$) plane
\citep[see ][ and references therein for details]{Papaderos2002}.
In doing so, we have assumed throughout an average $B-V$ colour of 0.5 mag 
for the systems studied here. 
An overlap between typical BCDs and our sample galaxies is also apparent
from the upper panel of the same figure where M$_{\rm LSB}$ is plotted against 
the central surface brightness $\mu_{\rm E,0}$ of the LSB host galaxy.
Similarly, the mean effective radius of r$_{\rm eff}$ (kpc) = 0.61$\pm$0.31 ($\sigma$=0.09) 
of the present sample matches closely typical values for BCDs, as derived 
from optical and near infrared surface photometry 
\citep[e.g., ][]{Cairos01a,Guseva01,Guseva03a,Doublier2002,Kai2003,Kai2005}.

The BCD nature of the emission-line galaxies discussed here is also reflected 
on the measured ratio of the effective radius r$_{\rm eff}$ to the scale length 
$\alpha$ of the LSB component. For a negligible starburst component on 
top of a purely exponential LSB host, one would expect the observed
effective radius to be approximately 1.7$\alpha$.
Conversely, a measured ratio r$_{\rm eff}/(1.7\alpha)<1$ 
is indicative of a central luminosity excess on top of
the exponential host galaxy, which for the objects under study 
can be plausibly attributed to the starburst emission.
As evident from Fig. \ref{r_eff}, the latter ratio is well below unity for all 
sample galaxies, irrespective of M$_{\rm LSB}$.
The horizontal dashed gray lines in the same figure indicate the expected ratio
r$_{\rm eff}/(1.7\alpha)$ when a centrally superposed star-forming source
approximated here with an exponential component 
with a scale length $\alpha$/4, is providing
respectively 0\%, 25\%, 50\% and 75\% of the total BCD light.

The mean ratio $<$r$_{\rm eff}/1.7\alpha>$=0.62$\pm$0.15 ($\sigma$=0.03) 
for our sample clearly indicates a substantial luminosity contribution from the starburst.
Indeed, from profile decomposition we infer the luminosity fraction
of the host galaxy to be LSB/total=0.52$\pm$0.17 ($\sigma=$0.03; col. 9 in 
Table \ref{tab:phot}) in good agreement with both the 
low r$_{\rm eff}/(1.7\alpha$) ratio and previous determinations 
of LSB/total in BCDs based on deep ground-based surface photometry 
\citep{Papaderos96b,SalzerNorton1999,Cairos01a,GdP2005}.

In summary, various lines of evidence indicate that most of the emission-line galaxies
included in the present study show photometric properties typical for BCDs. 
That is, in order of importance, i) an LSB host galaxy with M$_{\rm LSB}\ga$--17 $B$ mag,
characterized by a central surface brightness $\mu_B<23$ \sbb\ and a smaller exponential 
scale length than dIs/dEs of equal M$_{\rm LSB}$, ii) an optical LSB/total luminosity ratio
$\la$0.5 and iii) an effective radius r$_{\rm eff} \la 0.6$ kpc.
Additionally, our sample galaxies possess in their majority a smooth circular or elliptical 
LSB host, as is the case for $\sim$90\% of the local BCD population 
(the class of BCDs referred to as iE/nE in the Loose \& Thuan scheme
and numerous systems discovered in the UM survey).

Although spatially resolved colours are available for a few sample galaxies only, 
we consider it likely that 
the LSB host in most of the objects investigated here
is composed of a several Gyr old stellar population,
as is arguably the case for the overwhelming majority of nearby BCDs 
\citep[and references therein]{LT86,Kunth88,Papaderos96a,Cairos01a,Kai2003,GdP2005}.

\begin{figure*}[!ht]
\begin{picture}(17,5.5)
\put(-0.25,0){{\psfig{figure=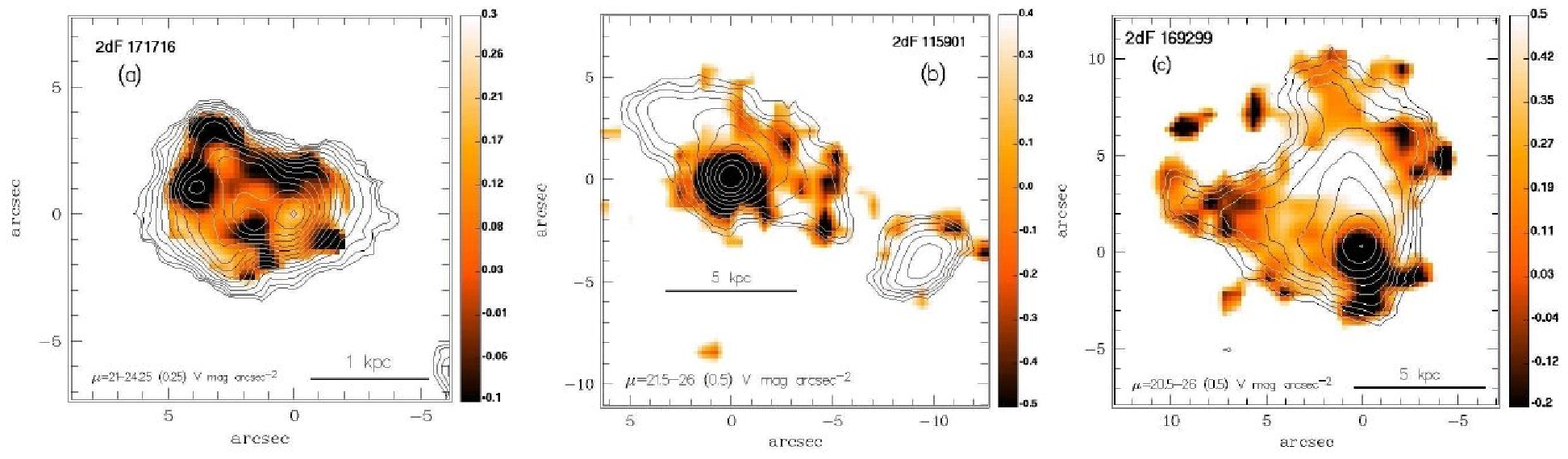,height=5.4cm,angle=0.,clip=}}}
\end{picture}
\caption{$V-I$ colour maps of three 2dFGRS galaxies displayed in the colour range 
indicated by the respective vertical bars. 
The surface brightness range of the superposed $V$ band contours and their step size 
are indicated at the lower-left of each panel.
North is to the top and east to the left.}
\label{vi}
\end{figure*}

Probable exceptions to this trend are the two most metal-poor galaxies 
in our sample, \object{2dF 171716} and \object{2dF 115901} (12+log(O/H)$<$7.6).
{\bb The $V-I$ profile of the first system shows a practically constant 
colour of $\sim$0.3 mag, suggesting a uniformly young stellar age 
out to its Holmberg radius (Fig. \ref{SBPs}).}
Such a colour constancy is {\bb very} uncommon among BCDs.
These systems exhibit almost invariably a strong colour 
gradient inside their 25 $B$ \sbb\ isophote as a typical signature of the increasing
line-of-sight contribution of the underlying old LSB host to the
measured intensity with increasing galactocentric radius
\citep[see, e.g., ][]{Papaderos96a,Papaderos2002,GdP2005}. 
The relatively small colour variation within \object{2dF 171716} with no 
clear distinction between high-surface brightness star-forming regions and 
the surrounding fainter envelope is {\bb also} apparent from the 
colour map in {\bb Fig. \ref{vi}a.
This was computed by subtracting the $I$ band image
from the $V$ band image after both images were co-registered,
smeared to the same angular resolution and calibrated.
The colour map is displayed in the range read off the vertical bar 
to the right of the image (--0.1\dots 0.3 mag) and adjusted such that 
its axis origin coincides with the brightest \h2\ region {\tt a}.
In order to better compare the broadband morphology with the color distribution 
we overlay on the $V-I$ map $V$ band contours between 21 and 24.25 \sbu\ 
in steps of 0.25 mag. 
It is apparent that the colour of \object{2dF 171716} does not gradually increase towards its
LSB periphery, as typically is the case in BCDs, but it shows an irregular pattern
with a mean value of $\sim$0.3 mag.
Clearly, deeper spectroscopic and photometric data are needed to decisively check
whether or not this emission-line galaxy is embedded within a more extended 
and faint stellar component with redder colours.}

Likewise, the evolutionary status of \object{2dF 115901} cannot be firmly constrained 
from the available data. This system displays a strong colour gradient 
inside $R^*=2$\arcsec\ and very blue $V-I$ colours ($\approx$--1 mag) in its
starburst region {\bb (see Figs. \ref{SBPs} \& \ref{vi}b).}
Whereas such a colour index is far too blue to be accounted for by any 
stellar population, it can readily be explained if optical broadband 
fluxes are severely contaminated be intense nebular forbidden and 
Balmer line emission.
This is indeed the case here, given that the measured equivalent 
width of the [O\,{\sc iii}]$\lambda$5007 attains values as large as 
$\approx$2000 \AA\ (cf. Table \ref{intensity}).
While for most star-forming galaxies at $z$$\approx$0 corrections 
of broad band colours for the contribution of ionized gas emission are small 
\citep[$\la$0.05 mag in $B-V$, see ][]{salzer89a}, 
there are several well-documented cases among BCDs where 
ample ionized gas emission falsifies colours by more than 
0.5 mag, not only in the close vicinity of young stellar clusters, 
but even on a global spatial scale. 
Most notable examples are the young BCD candidates I Zw 18 \citep{Papaderos2002} 
and SBS 0335-052\,E \citep{I97,Papaderos98}. 

As apparent from Fig. \ref{SBPs}, the radial colour profile of \object{2dF 115901} 
levels off for $R^*\ga 2\cdot r_{\rm eff}$ to a relatively blue mean value 
of $V-I=0.3 \pm 0.2$ mag, suggesting the presence of a predominantly 
young stellar population in its LSB host galaxy.
This fact, in connection with a subtle systematic colour variation 
between the redder northeastern and bluer southwestern part of the LSB component
calls for follow-up investigations of the dynamical 
and evolutionary status of this extremely metal-deficient BCD using deeper data.

A third system whose further investigation is likely of great interest is 
\object{2dF 169299} {\bb (Fig. \ref{vi}c)}. 
This irregular system displaying a $\Delta$-like morphology with two fanlike extentions 
towards NE, bears close resemblance to the BCD II\ Zw\ 40 \citep{SS1970}.
These two systems are also fairly similar with regard to their absolute $V$ magnitude
\citep[--17.6 mag for II\ Zw\ 40][as compared to --17.7 mag for \object{2dF 169299}]{Cairos01a}. 
They differ, however, in the exponential scale length of their LSB component 
which in the case of \object{2dF\ 169299} is twice as large as in II\,Zw\,40 
\citep[1.7 kpc as compared to $<$0.8 kpc; ][]{Cairos01a,GdP2005}. 
As evident from the colour map of the BCD the brightest star-forming region 
{\tt a} ($V-I \approx$--0.3) is surrounded by blue (0\dots 0.1 mag) 
patches along the northeastern extentions and at the position of region {\tt b}. 
The $V-I$ colour in the LSB component is comparatively blue, $\sim$0.4 mag,
possibly indicating a relatively young evolutionary 
state, provided that extended ionized gas emission does 
not significantly affect colours in the outskirts of this system.

Clearly, the sensitivity and photometric accuracy of the available broadband
data are not sufficient for a conclusive study of the evolutionary status 
of the three aforementioned BCDs. This kind of analysis was outside the scope and 
technical design of our observations, which have been solely optimized 
for the search of metal-deficient emission-line galaxies. 
In view of the present results, however, 
a closer investigation of the latter systems using deep follow-up imagery and 
spatially resolved spectroscopy is apparently of great interest.

\section{Summary \label{sum}}

We have extracted from the Two-Degree Field Galaxy Redshift Survey (2dFGRS) 100K 
Data Release a sample of 14 emission-line galaxies with a relatively strong 
[O {\sc iii}] $\lambda$4363 emission line which appear to be 
promising candidates for being new low-metallicity star-forming galaxies. 
Spectroscopic and photometric studies of this sample, and of 
7 Tololo and 2 UM additional galaxies were carried out using 
observations obtained with the ESO 3.6m telescope at La Silla.

For the majority of our sample galaxies, broad-band imaging reveals a 
low-surface brightness (LSB) host galaxy extending well beyond the 
star-forming component. 
As surface photometry in $V$ for all objects and in $I$ for some objects 
indicates, this LSB host is well approximated by an exponential fitting law 
and provides approximately half of the total $V$ emission, as is generally 
the case for blue compact dwarf (BCD) galaxies. 
Moreover, our photometric analysis indicates 
that the spectroscopically selected galaxies studied here are 
indistinguishable from BCDs regarding the structural properties 
(e.g., central surface brightness and exponential scale length) of their LSB host.
These facts establish a posteriori the BCD nature of our sample galaxies.

Based on longslit spectroscopy, we identify seven 2dF galaxies with an oxygen abundance 
12+log(O/H) $\la$ 7.8, and confirm previous claims that 
the star-forming systems \object{Tol 1304--353}, \object{Tol 2146--391}, \object{UM 559} and \object{UM 570} 
also have 12+log(O/H)$\leq$ 7.8. 
Additionally, we find that the distributions of our sample galaxies in 
the 12 +log(O/H) vs. log(Ne/O), log(Ar/O) and log(Fe/O) diagrams 
are similar to those of a comparison sample of 
$\sim$ 100 emission-line galaxies from \citet{ISGT04}.

Most importantly, we report the discovery of two 
galaxies, \object{2dF 115901} and \object{2dF 171716}, with an
oxygen abundance 12+log(O/H) $\sim$ 7.6 and $\sim$ 7.5, respectively.
This allows us to go another step forward in filling the 
gap between the most metal-deficient star-forming galaxies known
\object{SBS 0335-052 W\&E} (12+log(O/H)=7.1 and 7.3, respectively)
and \object{I Zw 18} (12+log(O/H)=7.2) and the majority of BCDs with a 
mean oxygen abundance 12+log(O/H)$>$7.6.

\color{black}
\begin{acknowledgements}
N.G.G. and Y.I.I. have been supported by DFG grants 436\ UKR 17/2/04 and 
436\ UKR 17/25/05. The research described in this paper was made possible in part 
by Award No. UP1-2551-KV-03 of the US Civilian Research \& Development Foundation 
for the Independent States of the Former Soviet Union (CRDF).
They are grateful for the hospitality of the G\"ottingen Observatory. 
T.X.T and Y.I.I have been partially supported by NSF grant AST-02-05785.
T.X.T. is grateful for a Sesquicentennial Fellowship from the 
University of Virginia. He thanks the hospitality of the Institut 
d'Astrophysique in Paris and of the Service d'Astrophysique at Saclay 
during his sabbatical leave.  
This research has made use of the NASA/IPAC Extragalactic Database (NED) which is operated by the 
Jet Propulsion Laboratory, California Institute of Technology, under contract with the National 
Aeronautics and Space Administration.
P.P. and K.G.N. would like to thank Linda Schmidtobreick, 
Jos\'e Cort\'es, Olivier Hainaut, George Hau, Erich Wenderoth 
and the whole ESO staff at the La Silla Observatory for their support.
\end{acknowledgements} 


\Online

\setcounter{footnote}{0}
\begin{table*} 
\caption{\label{decomp_res}Structural properties of the sample galaxies in the $V$ band.}
\label{tab:phot}
\begin{tabular}{lccccccccc}
\hline
\hline
Name  & $\mu_{\rm E,0}$ & m$_{\rm LSB}$ & m & r$_{\rm eff}$   & r$_{\rm Petr}$ &
log(r$_{80}$/r$_{20}$) &  r$_{\rm 25}$/r$_{\rm eff}$ & LSB/total & D \\
%
morph. type$\dag$ & $\alpha$        & M$_{\rm LSB}$ & M & $\mu_{\rm eff}$ & m$_{\rm Petr}$ &
1/$\eta$ &   fit range & ($V-I$)$_{\rm LSB}$ & $A_V$\\
(1)  &  (2)  & ( 3) &  (4) & (5) & (6) & (7) & (8) & (9) & (10) \upperspace\lowspace \\
\hline
2dF 336246 & 22.4$\pm$0.1  & 17.51 & 16.95 & 0.33  & 0.45  & 0.82 &
3.0 & 0.57 & 22.7\\
nE  & 0.41$\pm$0.02  &  --14.25   &  --14.83 & 21.31 & 17.50 & 0.38 & 
6\arcsec--16\farcs9 & 0.55$\pm$0.06 & 0.03\\
\hline
2dF 084585 & 21.7$\pm$0.8  & 18.42 & 17.9 & 0.75  & 1.27 & 0.60 & 0.49 &
 0.60 & 69.7\\
iI  & 0.62$\pm$0.15  &  --15.80     & --16.31  & 21.63 & 18.15 & 0.49 & 
4\farcs8--7\farcs6 & --- & 0.05\\
\hline
2dF 169299 & 22.4$\pm$0.4  & 18.57 & 18.22 & 1.87  & 3.42 & 0.79 & 
2.24 & 0.72 & 152.5\\
iI  & 1.69$\pm$0.2  &  --17.34   &  --17.70   & 22.23 & 18.54  & 0.46 &  
5\farcs6--10\farcs7 & 0.38$\pm$0.07 & 0.06\\
\hline
2dF 366973 & 21.47$\pm$0.14  & 17.77 & 17.40 & 0.22  & 0.39 & 0.84 & 
2.95 & 0.71 & 19.25\\
nE  & 0.2$\pm$0.01  &  --13.65   &  --14.02   & 21.28 & 17.78  & 0.50 & 
5\arcsec--12\farcs6 & 0.78$\pm$0.03 & 0.07\\
\hline
2dF 211354 & 21.7$\pm$0.35  & 18.9 & 17.7 & 0.59  & 0.93 & 0.54 & 
3.44 & 0.34 & 92\\
nE  & 0.66$\pm$0.06  &  --15.9   &  --17.2   & 20.26 & 17.99  & 0.54 & 
4\arcsec--8\farcs3 & -- & 0.12 \\
\hline
2dF 214945 & 22.05$\pm$0.3  & 18.27 & 17.37 & 1.24  & 1.72 & 0.64 & 
3.2 & 0.43 & 133.5\\
iI  & 1.47$\pm$0.13  &  --17.36   &  --18.26  & 20.78 & 17.86 & 0.54 &  
4\arcsec--12\arcsec & -- & 0.16\\
\hline
2dF 301921 & 21.91$\pm$0.14  & 18.49 & 18.18 & 0.36  & 0.47 & 0.73 & 
2.5 & 0.73  & 34.1\\
nE  & 0.32$\pm$0.02  &  --14.17   &  --14.48   & 21.87 & 18.74  & 0.55 &  
4\arcsec--10\arcsec & -- & 0.07\\
\hline
2dF 302461 & 22.48$\pm$0.36  & 18.6 & 17.89 & 0.29  & 0.45 & 0.65 & 
2.48 & 0.48 &  26.5\\
nE  & 0.31$\pm$0.04  &  --13.52   &  --14.23  & 21.67 & 18.29  & 0.61 & 
5\farcs7--10\farcs4 & --- & 0.1\\
\hline
Tol 1304--386 & 21.21$\pm$0.44  & 16.82 & 16.24 & 0.82  & 1.31 & 0.72 & 
3.3 & 0.57 & 52.6\\
iI  & 0.77$\pm$0.09  &  --16.79   &  --17.36   & 20.77 & 16.61  & 0.58 &  
10\farcs9--15\farcs3 &  --- & 0.1\\
\hline
Tol 1304--353 & 21.4$\pm$0.7  & 18.88 & 17.34 & 0.27  & 0.43 & 0.54 & 
4.0 & 0.24 & 52.1\\
nE  & 0.32$\pm$0.06  &  --14.71   &  --16.25   & 19.46 & 17.66  & 0.61 &  
3\farcs9--6\farcs3  & --- & 0.185\\
\hline
UM 559 & 21.8$\pm$0.8  & 16.28 & 15.64 & 0.54  & 0.87 & 0.51 & 
2.07 & 0.54 & 14.3\\
iE/iI  & 0.35$\pm$0.07  & --14.49   &  --15.14   & 22.09 & 23.35  & 0.84 & 
19\arcsec--26\arcsec & --- & 0.09\\
\hline
UM 570 & 21.96$\pm$0.55  & 19.98 & 18.10 & 0.34  & 0.64 & 0.49 & 
3.54 & 0.17 & 87.1\\
nE  & 0.42$\pm$0.08  &  --14.72   &  -16.6   & 19.58 & 18.29  & 0.52 & 
2\farcs8--3\farcs7 & --- & 0.10\\
\hline
Tol 1334--326 & 20.78$\pm$0.14  & 16.38 & 16.21 & 0.83 & 1.14 & 0.74 & 
3.2 & 0.8 & 45.8\\
iI,C  & 0.67$\pm$0.03  &  --16.92   &  --17.1   & 21.07 & 16.7 & 0.58 & 
9\arcsec--17\arcsec & --- & 0.17\\
\hline
2dF 323427 & 21.97$\pm$0.8  & 19.18 & 18.34 & 0.39  & 0.62 & 0.52 & 
2.3 & 0.44 & 47.2\\
iI,C  & 0.33$\pm$0.06  &  --14.19   & --15.03   & 21.48 & 18.59 & 0.87 & 
4\arcsec--6\farcs3 & --- & 0.16\\
\hline
Tol 1924--416 & 20.57$\pm$0.15  & 14.82 & 13.62 & 0.60  & 0.81 & 0.67 & 
7.03 & 0.33 & 37.8\\
iE  & 1.03$\pm$0.04  &  --18.06   &  --19.27   & 18.19 & 14.11  & 0.36 &  
14\farcs7--28\arcsec & --- & 0.29\\
\hline
Tol 2138--405 & 19.29$\pm$34  & 17.17 & 16.95 & 1.61  & 2.51 & 0.57 & 
3.6 & 0.82 & 224\\
nE/iI  & 1.15$\pm$0.1  & --19.58   &  --19.80   & 19.84 & 17.34  & 0.94 & 
3\farcs4--5\farcs5 & --- & 0.07\\
\hline
Tol 2146--391 & 19.79$\pm$0.36  & 18.30 & 17.38 & 0.51  & 0.75 & 0.52 & 
4.2 & 0.42 & 118\\
nE  & 0.45$\pm$0.04  &  --17.06   &  --17.99   & 19.12 & 17.73  & 0.70 &  
2\farcs7--5\arcsec & --- & 0.1\\
\hline
2dF 171716 & 20.96$\pm$0.6  & 18.37 & 17.89 & 0.46  & 0.74 & 0.47 & 
2.33 & 0.64 & 44.7\\
iI  & 0.28$\pm$0.04  &  --14.88   &  --15.36   & 21.52 & 18.14  & 1.01 &  
4\farcs6--8\arcsec & 0.2$\pm$0.06 & 0.08\\
\hline
2dF 116230 & 22.42$\pm$0.53  & 19.54 & 18.73 & 0.64  & 0.99 & 0.62 & 
2.61 & 0.47 & 94.2\\
nE  & 0.68$\pm$0.08  &  --15.33   &  --16.14   & 21.48 & 19.12 & 0.60 &  
3\farcs7--7\farcs1 & 0.72$\pm$0.07 & 0.07\\
\hline
2dF 181442 & 22.1$\pm$0.5  & 18.9 & 18.2 & 0.41  & 0.63 & 0.57 & 
2.6 & 0.51 & 48.2\\
nE  & 0.41$\pm$0.08  &  --14.5   &  --15.2   & 21.4 & 18.6  & 0.78 &  
4\farcs4--7\arcsec & --- & 0.074 \\
\hline
2dF 115901 & 22.04$\pm$0.08  & 19.36 & 18.92$\pm$0.04 & 0.84  & 0.73 & 0.82 &  3.44
 & 0.63 & 160.8\\
iI  & 1.07$\pm$0.01  &  --16.67  &  --17.11  & 21.09 & 19.79 & 0.54 &  
2\farcs6--6\farcs8 & 0.3$\pm$0.06 & 0.06 \upperspace \lowspace \\
%
\hline
\hline
\end{tabular}

\upperspace \noindent $\dag$: morphogical BCD type according to the
\citet{LT86} classification scheme.\\
\noindent $\mu_{\rm E,0}$, $\alpha$: central $V$ surface brightness (\sbb) and
exponential scale length (kpc) of the LSB host galaxy.\\ 
\noindent m$_{\rm LSB}$, M$_{\rm LSB}$: apparent and absolute $V$ magnitude of
the LSB host galaxy (mag).\\
\noindent  m, M: total apparent and absolute $V$ magnitude (mag) derived from
SBP integration.\\
\noindent r$_{\rm eff}$, $\mu_{\rm eff}$: $V$ band effective radius (kpc) and
mean surface brightness (\sbb) inside r$_{\rm eff}$.\\
\noindent r$_{\rm Petr}$, m$_{\rm Petr}$: Petrosian radius (kpc) and apparent
magnitude (mag) inside r$_{\rm Petr}$.\\
\noindent log(r$_{80}$/r$_{20}$): light concentration index based on the
ratio of the radius r$_{80}$ and r$_{20}$ enclosing, respectively, 80\% and 20\% of the total
$V$ light.\\
\noindent  $1/\eta$: S\'ersic exponent of the $V$ SBP. \\
\noindent r$_{\rm eff}$/r$_{\rm 25}$: ratio of the isophotal radius at 25 $V$
\sbb\ and the effective radius.\\
\noindent LSB/total, $(V-I)_{\rm host}$: luminosity fraction in the $V$ band and mean $V-I$ colour of the host galaxy.\\
\noindent $D$, $A_V$: adopted distance in Mpc and $V$ band Galactic absorption in mag.
\end{table*}
\setcounter{footnote}{0}
\small
\begin{longtable}{lcr|cr|cr|cr} 
\caption{\label{intensity}Emission line intensities}\\
\hline\hline
\endfirsthead
\caption{continued.}\\
\hline\hline
\endhead
\hline
\endfoot
 & \multicolumn{2}{c|}{\bf 336246}& \multicolumn{2}{c|}{\bf 084585}&
\multicolumn{2}{c|}{\bf 169299}&
\multicolumn{2}{c}{\bf 366973} \upperspace\lowspace \\ 
\cline{2-9}
Ion&$F$$^{\rm 1}$
&$EW$$^{\rm 2}$
&$F$$^{\rm 1}$
&$EW$$^{\rm 2}$
&$F$$^{\rm 1}$
&$EW$$^{\rm 2}$
&$F$$^{\rm 1}$
&$EW$$^{\rm 2}$  \upperspace\lowspace \\ \hline
3727\ [O {\sc ii}]             & 114.4$\pm$1.8  &  99.1$\pm$0.5 & 44.4$\pm$3.1   & 62.0$\pm$4.1  & 64.0$\pm$1.5   &  47.3$\pm$0.8 & 45.5$\pm$1.1   &  56.2$\pm$1.1  \\
3750\ H12                      &         ...      &         ...    &       ...       &       ...      &       ...       &       ...      &  1.9$\pm$0.5   &   2.4$\pm$0.7 \\
3771\ H11                      &         ...      &         ...    &       ...       &       ...      &       ...       &       ...      &  2.6$\pm$0.6   &   3.4$\pm$0.8 \\
3798\ H10                      & 3.6$\pm$0.6  &   3.2$\pm$0.5 &       ...       &       ...      &  4.6$\pm$1.4   &   3.4$\pm$1.0 &  4.3$\pm$0.7   &   5.8$\pm$0.9 \\
3819\ He {\sc i}               & 1.6$\pm$0.6  &   1.4$\pm$0.5 &       ...       &       ...      &       ...       &      ...       &  0.7$\pm$0.6   &   1.0$\pm$0.8 \\
3835\ H9                       & 4.4$\pm$0.4   &   3.8$\pm$0.3 & 7.3$\pm$2.0   & 12.3$\pm$3.3 &  5.2$\pm$1.1   &   3.8$\pm$0.8 &  6.2$\pm$0.6   &   8.6$\pm$0.8 \\
\hline\hline
\end{longtable}


\small
\begin{longtable}{lcccccccc} 
\caption{\label{heavy}Heavy element abundances}\\
\hline\hline
Property& {\bf 336246} & {\bf 084585} & {\bf 169299} & {\bf 366973} & {\bf
  211354} & {\bf 214945} & {\bf 301921} & {\bf 302461} \upperspace\lowspace  \\ 
\hline
\endfirsthead
\caption{continued.}\\
\hline\hline
\endhead
\hline
\endfoot
$t_e$(O {\sc iii})(K)$^{\rm 5}$                     &1.53$\pm$0.02     &1.90$\pm$0.08     &1.75$\pm$0.04     &1.66$\pm$0.04     &1.41$\pm$0.02     &1.11$\pm$0.03     &1.69$\pm$0.06     &1.55$\pm$0.04        \\ 
$t_e$(O {\sc ii})(K)$^{\rm 5}$                       &1.39$\pm$0.02     &1.54$\pm$0.06     &1.48$\pm$0.03     &1.45$\pm$0.03     &1.33$\pm$0.04     &1.16$\pm$0.03     &1.46$\pm$0.05     &1.40$\pm$0.04             \\ \\
O$^+$/H$^+$($\times$10$^5$)         &1.29$\pm$0.06    &0.36$\pm$0.04    &0.57$\pm$0.04    &0.49$\pm$0.03    &1.42$\pm$0.07    &4.30$\pm$0.37    &1.13$\pm$0.10    &1.15$\pm$0.09           \\
O$^{++}$/H$^+$($\times$10$^5$)      &5.14$\pm$0.20    &4.26$\pm$0.43    &4.12$\pm$0.24    &5.41$\pm$0.31    &7.57$\pm$0.36    &12.82$\pm$1.13    &4.62$\pm$0.38    &4.96$\pm$0.35           \\
O$^{+3}$/H$^+$($\times$10$^5$)      &0.08$\pm$0.02    &...            &0.15$\pm$0.03    &0.10$\pm$0.07    &...            &...             &...            &...                   \\
O/H($\times$10$^5$)                 &6.52$\pm$0.21    &4.62$\pm$0.43    &4.84$\pm$0.25    &6.00$\pm$0.32    &8.99$\pm$0.37    &17.11$\pm$1.19    &5.75$\pm$0.39    &6.11$\pm$0.36           \\
12 + log(O/H)                       &7.81$\pm$0.01    &7.66$\pm$0.04    &7.68$\pm$0.02    &7.78$\pm$0.02    &7.95$\pm$0.02    &8.23$\pm$0.03    &7.76$\pm$0.03    &7.79$\pm$0.02           \\
\hline\hline
\end{longtable}


\begin{thebibliography}{}

\bibitem[Asplund et al. (2005)]{Asplund04} Asplund, M., Grevesse, N., \&
Sauval, A. J. 2005, in {\sl Cosmic Abundances as Records of Stellar Evolution 
and Nucleosynthesis}, ASP Conf. Ser. 336, 25

\bibitem[Bicker et al. (2004)]{Bicker2004} Bicker, J., 
Fritze-v.~Alvensleben, U., M\"oller, C.~S., Fricke, K.~J. 2004, \aap, 413, 37

\bibitem[Binggeli \& Cameron (1991)]{binggeli91} Binggeli, B., 
\& Cameron, L. M. 1991, \aap, 252, 27

\bibitem[Binggeli \& Cameron (1993)]{binggeli93} Binggeli, B., 
\& Cameron, L. M. 1993, \aaps, 98, 297

\bibitem[Bothun et al. (1991)]{bothun91} Bothun, G. D., Impey, C. D., 
\& Malin, D. F. 1991, \apj, 376, 404

\bibitem[Cair\'os et al. (2001a)]{Cairos01a} Cair\'os, L.~M., 
V\'ilchez, J.~M., Gonz\'alez P\'erez, J.~N., Iglesias-P\'aramo, J., 
Caon, N., 2001a, \apjs 133, 321

\bibitem[Cair\'os et al. (2001b)]{Cairos01b} Cair\'os, L.~M., Caon, N., 
V\'ilchez, J.~M., Gonz\'alez P\'erez, J.~N., Mu\~noz-Tu\~n\'on, C. 2001b, 
\apjs, 133, 321

\bibitem[Caldwell \& Bothun (1987)]{caldwell87} Caldwell, N., \& Bothun, G. D.
 1987, \aj, 94, 1126

\bibitem[Campbell (1992)]{Campbell92} Campbell, A. 1992, \apj, 401, 157

\bibitem[Campbell et al. (1986)]{Campbell86} Campbell, A. W., Terlevich, R., 
\& Melnick, J. 1986, \mnras, 223, 811

\bibitem[Carignan \& Beaulieu (1989)]{carignan89} Carignan, C., 
\& Beaulieu, S. 1989, \apj, 347, 760

\bibitem[Colless et al. (2001)]{C01} Colless, M., Dalton, G., Maddox, S., 
Sutherland, W., Norberg, P. et al. 2001, \mnras, 328, 1039

\bibitem[Doublier et al. (2002)]{Doublier2002} Doublier, V., Caulet, A., 
Comte, G. 2002, \aap, 367, 33      

\bibitem[Drinkwater \& Hardy (1991)]{drinkwater91} Drinkwater, M., 
\& Hardy, E. 1991, \aj, 101, 94

\bibitem[Fioc \& Rocca-Volmerange (1997)]{FRV97} Fioc, M., \& 
Rocca-Volmerange, B. 1997, \aap, 326, 950

\bibitem[Fricke et al. (2001)]{Fricke01} Fricke, K.J., Izotov, Y.I., 
Papaderos, P., Guseva, N.G., Thuan, T.X. 2001, \aj, 121, 169

\bibitem[Gil de Paz \& Madore (2005)]{GdP2005} Gil de Paz, A. \& Madore, B.F. 
2005, \apjs, 156, 345

\bibitem[Gil de Paz et al. (2003)]{GdP2003} Gil de Paz, A., Madore, B.F., 
\& Pevunova, O. 2003, \apjs, 147, 29

\bibitem[Guseva et al. (2001)]{Guseva01} Guseva, N.G., Izotov, Y.I., 
Papaderos, P., Chaffee, F.H., Foltz, C.B., Green, R.F., Thuan, T.X., 
Fricke, K.J., Noeske, K.G. 2001, \aap, 378, 756

\bibitem[Guseva et al. (2003)]{Guseva03a} Guseva, N.G., Papaderos, P., 
Izotov, Y.I., Green, R.F., Fricke, K.J., Thuan, T.X., Noeske, K.G. 
2003, \aap, 407, 75

\bibitem[Guseva et al. (2004)]{Guseva04} Guseva, N. G., Papaderos, P., 
Izotov, Y. I., Noeske, K. G., \& Fricke, K. J. 2004, \aap, 421, 519

\bibitem[Hidalgo-G\'amez et al. (2003)]{Hid03} Hidalgo-G\'amez, A. M., 
S\'anchez-Salcedo, F. J., \& Olofsson, K. 2003, \aap, 399, 63

\bibitem[Hopp \& Schulte-Ladbeck (1991)]{Hopp91} Hopp, U., \& 
Schulte-Ladbeck, R. E. 1991, \aap, 248, 1

\bibitem[Izotov \& Thuan (1998)]{IT98} Izotov, Y. I., \& Thuan, T. X. 
1998, \apj, 497, 227 

\bibitem[Izotov \& Thuan (2004a)]{IT04a} Izotov, Y. I., \& Thuan, T. X. 
2004a, \apj, 602, 200 

\bibitem[Izotov \& Thuan (2004b)]{IT04b} Izotov, Y. I., \& Thuan, T. X. 
2004b, \apj, 616, 768 

\bibitem[Izotov et al. (1994)]{ITL94} Izotov, Y. I., Thuan, T. X., 
\& Lipovetsky, V. A. 1994, \apj, 435, 647  

\bibitem[Izotov et al. (1997a)]{ITL97} Izotov, Y. I., Thuan, T. X., 
\& Lipovetsky, V. A. 1997a, \apjs, 108, 1

\bibitem[Izotov et al. (1997b)]{I97} Izotov, Y. I., Lipovetsky, V. A., 
Chaffee, F. H., Foltz, C. B., Guseva, N. G., Kniazev, A. Y. 1997b, 
\apj, 476, 698

\bibitem[Izotov et al. (1999)]{ICF99} Izotov, Y. I., Chaffee, F. H., 
Foltz, C. B., et al. 1999, \apj, 527, 757

\bibitem[Izotov et al. (2001)]{ICF01} Izotov, Y. I., Chaffee, F. H., 
\& Schaerer, D. 2001, \aap, 378, L45

\bibitem[Izotov et al. (2004a)]{Iz04} Izotov, Y. I., Noeske, K. G., 
Guseva, N. G., Papaderos, P., Thuan, T. X., \& Fricke, K. J. 
2004a, \aap, 415, L27

\bibitem[Izotov et al. (2004b)]{ISGT04} Izotov, Y. I., Stasi\'nska, G., 
Guseva, N. G., Thuan, T. X. 2004b, \aap, 415, 87 

\bibitem[Izotov et al. (2004c)]{Izotov2004} Izotov, Y.I., Noeske, K.G., 
Guseva, N.G., Papaderos, P., Thuan, T.X. \& Fricke, K.J. 
2004c, \aap, 415, 27L

\bibitem[Izotov et al. (2005)]{ITG05} Izotov, Y. I., Thuan, T. X., 
\& Guseva, N. G. 2005, \apj, 632, 210

\bibitem[Izotov et al. (2006)]{I06} Izotov, Y. I., Stasi\'nska, G., 
Meynet, G., Guseva, N. G., \& Thuan, T. X. 2006, \aap, 448, 955  

\bibitem[Kehrig et al. (2004)]{kehrig04} Kehrig, K., Telles, E., 
\& Cuisinier, F. 2004, \aj, 128, 1141

\bibitem[Kobulnicky \& Skillman (1996)]{kobul96} Kobulnicky, H. A., 
\& Skillman, E. D. 1996, \apj, 471, 211

\bibitem[Kunth et al. (1988)]{Kunth88} Kunth, D., Maurogordato, S., 
\& Vigroux, L. 1988, \aap, 204, 10

\bibitem[Kraan-Korteweg (1986)]{K86} Kraan-Korteweg, R. C. 1986, 
\aaps, 66, 255

\bibitem[Landolt (1992)]{Lan92} Landolt, A.U. 1992, \aj, 104, 340

\bibitem[Lauberts \&  Valentijn (1989)]{LV89} Lauberts, A. \&  Valentijn, E.A.
1989, {\sl The Surface Photometry Catalogue of the ESO-Uppsala Galaxies, ESO}

\bibitem[Loose \& Thuan (1986)]{LT86} Loose, H. H. \& Thuan, T. X. 1986, 
Star Forming Dwarf Galaxies and Related Objects, Editions Frontieres, 73 

\bibitem[Maddox et al. (1990)]{Maddox90} Maddox, S.J., Sutherland, W.J., 
Efstathiou, G., \& Loveday, J. 1990, \mnras, 243, 692

\bibitem[Marconi et al. (1994)]{Marconi94} Marconi, G., Matteucci., F., 
\& Tosi, M. 1994, \mnras, 270, 35 

\bibitem[Marlowe et al. (1997)]{Mar97} Marlowe, A. T., Meurer, G. R., 
Heckman, T. M., \& Schommer, R. 1997, \apjs, 112, 285

\bibitem[Masegosa et al. (1994)]{Masegosa94} Masegosa, J., Moles, M., 
\& Campos-Aguilar, A. 1994, \apj, 420, 576

\bibitem[Noeske (1999)]{noeske99} Noeske, K. G. 1999, Diploma Thesis 

\bibitem[Noeske et al. (2000)]{Kai00} Noeske, K. G., Guseva, N. G., Fricke, K. J., et al. 2000, \aap, 361, 33

\bibitem[Noeske et al. (2003)]{Kai2003} Noeske, K.G., Papaderos, P., 
Cair\'os, L.M., Fricke, K.J. 2003, \aap, 410, 481

\bibitem[Noeske et al. (2005)]{Kai2005} Noeske, K.G., Papaderos, P., 
Cair\'os, L.M., Fricke, K.J. 2005, \aap, 429, 115

\bibitem[Pagel et al. (1986)]{Pagel86} Pagel, B.E.J., Terlevich, R.J., 
\& Melnick, J. 1986, \mnras, 98, 1005

\bibitem[Pagel et al. (1992)]{Pagel92} Pagel, B.E.J., Simonson, E.A., 
Terlevich, R.J. \& Edmunds, M.G. 1992, \mnras, 255, 325

\bibitem[Papaderos et al. (1996a)]{Papaderos96a} Papaderos, P., Loose, H.-H., 
Thuan, T.X. \& Fricke, K.J. 1996a, \aaps, 120, 207 

\bibitem[Papaderos et al. (1996b)]{Papaderos96b} Papaderos, P., Loose, H.-H., 
Fricke, K.J. \& Thuan, T. X. 1996b, \aap, 314, 59 

\bibitem[Papaderos et al. (1998)]{Papaderos98} Papaderos, P., Izotov, Y.I., 
Fricke, K.J., Thuan, T.X., Guseva, N.G. 1998, \aap, 338, 43

\bibitem[Papaderos et al. (2002)]{Papaderos2002} Papaderos, P., Izotov, Y.I., 
Thuan, T.X., Noeske, K.G., Fricke, K.J., Guseva, N.G., Green, R.F. 
2002, \aap, 292, 461

\bibitem[Patterson \& Thuan (1996)]{PT96} Patterson, R. J., \& Thuan, T. X. 
1996, \apjs, 107, 103

\bibitem[Peimpert \& Torres-Peimpert (1974)]{PeimpertTP1974}Peimpert, M. \& Torres-Peimpert, S. 1974, ApJ, 193, 327

\bibitem[Peimpert \& Torres-Peimpert (1976)]{PeimpertTP1976}Peimpert, M. \& Torres-Peimpert, S. 1976, ApJ, 203, 581

\bibitem[Petrosian (1976)]{Petrosian76} Petrosian, V. 1976, \apj, 210, L53

\bibitem[Pilyugin (2000)]{Pil2000} Pilyugin, L. S. 2000, \aap, 362, 325

\bibitem[Pustilnik et al. (2004)]{PPK04} Pustilnik, S. A., Pramskij, A. G., 
\& Kniazev, A. Y. 2004, \aap, 425, 51

\bibitem[Richer \& McCall (1995)]{richer95} Richer, M. G., \& McCall, M. L. 
1995, \apj, 445, 642

\bibitem[Salzer et al. (1989a)]{salzer89a} Salzer, J. J., MacAlpine, G. M., 
\& Boroson, T. A. 1989a, \apjs, 70, 447

\bibitem[Salzer et al. (1989b)]{salzer89b} Salzer, J. J., MacAlpine, G. M., 
\& Boroson, T. A. 1989b, \apjs, 70, 479

\bibitem[Salzer \& Norton (1999)]{SalzerNorton1999} Salzer, J.J., \& 
Norton, S.A. 1999, in the {\sl Low Surface Brightness Universe}, 
ASP Conf. Ser. 170, 253

\bibitem[Sargent \& Searle (1970)]{SS1970} Sargent, W.L.W., \& Searle, L. 
1970, \apj, 152, L155

\bibitem[Schlegel et al. (1989)]{Schlegel89} Schlegel, D.J., Finkbeiner, D.P.,
\& Davis, M. 1989, \apj, 500, 525

\bibitem[Searle \& Sargent (1972)]{SS1972} Searle, L., \& Sargent, W.L.W. 
1972, \apj, 173, 25

\bibitem[Stasi\'nska \& Izotov (2003)]{SI03} Stasi\'nska, G. \& Izotov, Y. I.
2003, \aap, 397, 71

\bibitem[Takamiya (1999)]{Takamiya99} Takamiya, M. 1999, \apjs, 122, 109

\bibitem[Telles \& Terlevich (1995)]{telles95} Telles, E., \& Terlevich, R. 
1995, \mnras, 275, 1

\bibitem[Terlevich et al. (1991)]{terl91} Terlevich, R., Melnick, J., 
Masegosa, J., Moles, M., \& Copetti, M.V.F. 1991, \aaps, 91, 285

\bibitem[Thuan \& Izotov (2005)]{TI05} Thuan, T. X., \& Izotov, Y. I. 
2005, \apjs, 161, 240

\bibitem[Thuan et al. (1995)]{til95} Thuan, T. X., Izotov, Y. I., 
\& Lipovetsky, V. A. 1995, \apj, 445, 108

\bibitem[van Zee (2000)]{vZ00} van Zee, L. 2000, \aj, 119, 2757

\bibitem[van Zee et al. (2004)]{vanZee2004} van Zee, L., Skillman, E.D., 
\& Haynes, M.P., 2004, \aj, 128, 121

\bibitem[Vigroux et al. (1986)]{vigroux86} Vigroux, L., Thuan, T. X., 
Vader, J. P., \& Lachi\`eze--Rey, M. 1986, \aj, 91, 70

\end{thebibliography}
\end{document}